\documentclass{aa}
\usepackage{natbib}
\bibpunct{(}{)}{;}{a}{}{,} 

\usepackage[english]{babel}           
\usepackage{graphicx,epsfig}
\usepackage{txfonts}
\usepackage{mathrsfs}
\usepackage{lscape}
\usepackage{eucal}
\usepackage{pifont}
\usepackage{rotating}
\usepackage{multirow}
\usepackage{tikz}
\usepackage[colorlinks=true,citecolor=blue,linkcolor=blue]{hyperref}
\usepackage{longtable}
\usepackage[fleqn]{amsmath}
\usepackage{breqn}

\def \cc    {\ifmmode{\,{\rm cm}^{-3}}\else{$\,{\rm cm}^{-3}$}\fi}
\def \cq    {\ifmmode{\,{\rm cm}^{-2}}\else{$\,{\rm cm}^{-2}$}\fi}
\def \mic   {\ifmmode{\,\mu{\rm m}}\else{$\mu$m}\fi}
\def \eccs  {\ifmmode{\,{\rm erg}\,{\rm cm}^{-3} {\rm s}^{-1}}\else{$\,{\rm erg}\,{\rm cm}^{-3} {\rm s}^{-1}$}\fi}
\def \ecc   {\ifmmode{\,{\rm erg}\,{\rm cm}^{-3}}\else{$\,{\rm erg}\,{\rm cm}^{-3}$}\fi}
\def \ecqs  {\ifmmode{\,{\rm erg}\,{\rm cm}^{-2}\,{\rm s}^{-1}\,{\rm 
             sr}^{-1}}\else{$\,{\rm erg}\,{\rm cm}^{-2}\,{\rm s}^{-1}\,{\rm sr}^{-1}$}\fi}
\def \ecss  {\ifmmode{\,{\rm erg}\,{\rm cm}^{-2}\,{\rm s}^{-1}}\else{$\,{\rm erg}\,{\rm cm}^{-2}\,{\rm s}^{-1}$}\fi}
\def \deg   {\ifmmode{^{\circ}}\else{$^{\circ}$}\fi} 
\def \pc    {\ifmmode{\,{\rm pc}}\else{$\,{\rm pc}$}\fi} 
\def \kms   {\ifmmode{\,{\rm km}\,{\rm s}^{-1}}\else{km\,\,s$^{-1}$}\fi} 
\def \kmspc {\ifmmode{\,{\rm km}\,{\rm s}^{-1}\,{\rm pc}^{-1}}\else{km s$^{-1}$ pc$^{-1}$}\fi} 
\def \MJysr {\ifmmode{\,{\rm MJy\,sr}^{-1}}\else{$\,{\rm MJy\,sr}^{-1}$}\fi} 
\def \Kkms  {\ifmmode{\,{\rm K\,km\,s}^{-1}}\else{$\,{\rm K\,km\,s}^{-1}$}\fi}
\def \epso{\ifmmode{\overline{\varepsilon}_{\rm obs}}\else{$\overline{\varepsilon}_{\rm obs}$}\fi}
\def \utM{\ifmmode{u_{\theta,{\rm M}}}\else{$u_{\theta,{\rm M}}$}\fi}
\def \urM{\ifmmode{u_{r,{\rm M}}}\else{$u_{r,{\rm M}}$}\fi}
\def \twCO{\ifmmode{\rm ^{12}CO}\else{$\rm^{12}CO$}\fi} 
\def \thCO{\ifmmode{\rm ^{13}CO}\else{$\rm^{13}CO$}\fi} 
\def \CeiO{\ifmmode{\rm C^{18}O}\else{$\rm C^{18}O$}\fi} 
\def \twCN{\ifmmode{\rm ^{12}CN}\else{$\rm^{12}CN$}\fi} 
\def \thCN{\ifmmode{\rm ^{13}CN}\else{$\rm^{13}CN$}\fi} 
\def \HdCO{\ifmmode{\rm H_{2}CO}\else{$\rm H_{2}CO$}\fi} 
\def \twHdCO{\ifmmode{\rm ^{12}H_{2}CO}\else{$\rm^{12}H_{2}CO$}\fi} 
\def \thHdCO{\ifmmode{\rm ^{13}H_{2}CO}\else{$\rm^{13}H_{2}CO$}\fi} 
\def \twC{\ifmmode{\rm ^{12}C}\else{$\rm^{12}C$}\fi} 
\def \thC{\ifmmode{\rm ^{13}C}\else{$\rm^{13}C$}\fi} 
\def \Hp{\ifmmode{\rm H^+}\else{$\rm H^+$}\fi} 
\def \Cp{\ifmmode{\rm C^+}\else{$\rm C^+$}\fi} 
\def \Sp{\ifmmode{\rm S^+}\else{$\rm S^+$}\fi} 
\def \Op{\ifmmode{\rm O^+}\else{$\rm O^+$}\fi} 
\def \CFp{\ifmmode{\rm CF^+}\else{$\rm CF^+$}\fi}
\def \CHp{\ifmmode{\rm CH^+}\else{$\rm CH^+$}\fi}
\def \CHdp{\ifmmode{\rm CH_2^+}\else{$\rm CH_2^+$}\fi}
\def \CHtp{\ifmmode{\rm CH_3^+}\else{$\rm CH_3^+$}\fi} 
\def \SHp{\ifmmode{\rm SH^+}\else{$\rm SH^+$}\fi}
\def \SHdp{\ifmmode{\rm SH_2^+}\else{$\rm SH_2^+$}\fi}
\def \SHtp{\ifmmode{\rm SH_3^+}\else{$\rm SH_3^+$}\fi}
\def \twCHp{\ifmmode{\rm ^{12}CH^+}\else{$\rm^{12}CH^+$}\fi}
\def \thCHp{\ifmmode{\rm ^{13}CH^+}\else{$\rm^{13}CH^+$}\fi}
\def \CtH{\ifmmode{\rm C_2H}\else{$\rm C_2H$}\fi} 
\def \CthHt{\ifmmode{\rm C_3H_2}\else{$\rm C_3H_2$}\fi} 
\def \Htp{\ifmmode{\rm H_3^+}\else{$\rm H_3^+$}\fi} 
\def \COp{\ifmmode{\rm CO^+}\else{$\rm CO^+$}\fi} 
\def \HCOp{\ifmmode{\rm HCO^+}\else{$\rm HCO^+$}\fi} 
\def \HtOp{\ifmmode{\rm H_3O^+}\else{$\rm H_3O^+$}\fi} 
\def \HCfiN{\ifmmode{\rm HC_5N}\else{$\rm HC_5N$}\fi} 
\def \wat{\ifmmode{\rm H_2O}\else{$\rm H_2O$}\fi} 
\def \HdO{\ifmmode{\rm H_2O}\else{$\rm H_2O$}\fi} 
\def \OHp{\ifmmode{\rm OH^+}\else{$\rm OH^+$}\fi} 
\def \HdOp{\ifmmode{\rm H_2O^+}\else{$\rm H_2O^+$}\fi} 
\def \HtOp{\ifmmode{\rm H_3O^+}\else{$\rm H_3O^+$}\fi} 
\def \NHd{\ifmmode{\rm NH_2}\else{$\rm NH_2$}\fi} 
\def \NHtrois{\ifmmode{\rm NH_3}\else{$\rm NH_3$}\fi} 
\def \oxy{\ifmmode{\rm O_2}\else{$\rm O_2$}\fi} 
\def \HH{\ifmmode{\rm H_2}\else{$\rm H_2$}\fi}
\def \Jone{\ifmmode{\rm {(J=1--0)}}\else{{(J=1--0)}}\fi} 
\def \Jtwo{\ifmmode{\rm {(J=2--1)}}\else{{(J=2--1)}}\fi} 
\def \Jthr{\ifmmode{\rm {(J=3--2)}}\else{{(J=3--2)}}\fi} 
\def \Jfou{\ifmmode{\rm {(J=4--3)}}\else{{(J=4--3)}}\fi} 
\def \Jfiv{\ifmmode{\rm {J=4--3}}\else{{J=4--3}}\fi} 
\def \Ta{\ifmmode{\rm T_A}\else{$\rm T_A$}\fi} 
\def \Tas{\ifmmode{\rm T_A^*}\else{$\rm T_A^*$}\fi} 
\def \Tmb{\ifmmode{\rm T_{mb}}\else{$\rm T_{mb}$}\fi} 
\def \Tr{\ifmmode{\rm T_r}\else{$\rm T_r$}\fi} 
\def \Trs{\ifmmode{\rm T_r^*}\else{$\rm T_r^*$}\fi}
\def \NHt{\ifmmode{N_{\rm H}}\else{$N_{\rm H}$}\fi}
\def \NH{\ifmmode{N({\rm H})}\else{$N({\rm H})$}\fi}
\def \NH2{\ifmmode{N({\rm H}_2)}\else{$N({\rm H}_2)$}\fi}
\def \NCH{\ifmmode{N({\rm CH})}\else{$N({\rm CH})$}\fi}
\def \NHF{\ifmmode{N({\rm HF})}\else{$N({\rm HF})$}\fi}
\def \dens{\ifmmode{n_{\rm H}}\else{$n_{\rm H}$}\fi}
\def \nCO{\ifmmode{n({\rm CO})}\else{$n({\rm CO})$}\fi}
\def \nHF{\ifmmode{n({\rm HF})}\else{$n({\rm HF})$}\fi}
\def \nH2{\ifmmode{n({\rm H}_2)}\else{$n({\rm H}_2)$}\fi}

\defcitealias{Bellomi2020}{Paper~I}

\begin{document}

\title{3D chemical structure of the diffuse turbulent ISM}
\subtitle{II - Origin of \CHp, new solution to an 80 years mystery}

\author{
  B. Godard            \inst{\ref{lerma}, \ref{ENS}}, 
  G. Pineau des Forêts \inst{\ref{IAS},  \ref{lerma}}, 
  P. Hennebelle        \inst{\ref{CEA}},
  E. Bellomi           \inst{\ref{Harvard}}, \and
  V. Valdivia          \inst{\ref{Nagoya}}
}

\institute{
Observatoire de Paris, Université PSL, Sorbonne Université, LERMA, 75014 Paris, France
\label{lerma}
\and
Laboratoire de Physique de l’Ecole Normale Supérieure, ENS, Université PSL, CNRS, Sorbonne Université, Université de Paris, F-75005 Paris, France
\label{ENS}
\and
Université Paris-Saclay, CNRS, Institut d’Astrophysique Spatiale, 91405, Orsay, France
\label{IAS}
\and
Laboratoire AIM, CEA/IRFU, CNRS/INSU, Universit\'e Paris Diderot, CEA-Saclay, 91191 Gif-Sur-Yvette, France 
\label{CEA}
\and
Harvard-Smithsonian Center for Astrophysics, 60 Garden Street, Cambridge, MA, USA
\label{Harvard}
\and
Department of Physics, Graduate School of Science, Nagoya University, Furo-cho, Chikusa-ku, Nagoya 464-8602, Japan
\label{Nagoya}
}

 \date{Received 29 April 2022 / Accepted 17 September 2022}

\abstract{}
{The large abundances of \CHp\ in the diffuse interstellar medium (ISM) are a long standing issue of our understanding of the thermodynamical and chemical states of the gas. We investigate, here, the formation of \CHp\ in turbulent and multiphase environments, where the heating of the gas is almost solely driven by the photoelectric effect.}
{The diffuse ISM is simulated using the magnetohydrodynamic (MHD) code RAMSES which self-consistently computes the dynamical and thermal evolution of the gas along with the time-dependent evolutions of the abundances of \Hp, H, and \HH. The rest of the chemistry, including the abundance of \CHp, is computed in post-processing, at equilibrium, under the constraint of out-of-equilibrium of \Hp, H, and \HH. The comparison with the observations is performed taking into account an often neglected, yet paramount, piece of information, namely the length of the intercepted diffuse matter along the observed lines of sight.}
{The quasi totality of the mass of \CHp\ originates from the unstable gas, in environments where the kinetic temperature is larger than $600$ K, the density ranges between 0.6 and 10 \cc, the electronic fraction ranges between $3 \times 10^{-4}$ and $6 \times 10^{-3}$, and the molecular fraction is smaller than $0.4$. Its formation is driven by warm and out-of-equilibrium \HH\ initially formed in the cold neutral medium (CNM) and injected in more diffuse environments and even the warm neutral medium (WNM) through a combination of advection and thermal instability. 
The simulation which displays the tightest agreement with the HI-to-\HH\ transition and the thermal pressure distribution observed in the Solar Neighborhood is found to naturally reproduce the  observed abundances of \CHp, the dispersion of observations, the probability of occurrence of most of the lines of sight, the fraction of non-detections of \CHp, and the distribution of its line profiles. The amount of \CHp\ and the statistical properties of the simulated lines of sight are set by the fraction of unstable gas rich in \HH\ which is controlled, on Galactic scales, by the mean density of the diffuse ISM (or, equivalently, its total mass), the amplitude of the mean UV radiation field, and the strength of the turbulent forcing.}
{This work offers a new and natural solution to an 80 years old chemical riddle. The almost ubiquitous presence of \CHp\ in the diffuse ISM likely results from the exchanges of matter between the CNM and the WNM induced by the combination of turbulent advection and thermal instability, without the need to invoke ambipolar diffusion or regions of intermittent turbulent dissipation. Through two phase turbulent mixing, \CHp\ might thus be a tracer of the \HH\ mass loss rate of CNM clouds.}

\keywords{ISM: structure - ISM: molecules - ISM: kinematics and dynamics - ISM: clouds - methods: numerical - methods: statistical}

\authorrunning{B. Godard et al.}
\titlerunning{production of \CHp\ through two phase turbulent mixing} 
\maketitle

\section{Introduction}

The methylidyne cation \CHp\ is among the molecules the most frequently seen in the diffuse interstellar medium. Since its first detection in 1941 by \citet{Douglas1941}, \CHp\ has been observed in absorption along a great variety of diffuse Galactic lines of sight, first with optical facilities (see Appendix \ref{AppendObs}) and later on with the {\it Herschel Space Telescope} (e.g., \citealt{Falgarone2010,Godard2012}). Surprisingly, very few lines of sight are dark in \CHp. The presence of \CHp\ in the diffuse ISM is so ubiquitous that this molecular ion is now frequently detected in external galaxies, including nearby objects (e.g., \citealt{Rangwala2011,Spinoglio2012,Ritchey2015}) and Starburst galaxies at high redshift (e.g., \citealt{Falgarone2017}).

The large abundances and the remarkable coverage of \CHp\ raise a chemical conundrum. Easily destroyed by collisions with H, \HH, and e$^-$, or by photodissociation, \CHp\ requires an efficient formation pathway. The only reaction capable of balancing its fast destruction is $\Cp + \HH \rightarrow \CHp + {\rm H}$, a highly endothermic chemical process ($\Delta E/k \sim 4640$ K, where $k$ is the Boltzmann constant). On the one side, the formation of \CHp\ can only proceed at high effective temperature (see Eq. 2.1 of \citealt{Pineau-des-Forets1986a}). On the other side, it requires molecular hydrogen, which is mainly formed in cold environments.

One possible solution to this issue is the release of suprathermal energy induced by the intermittent dissipation of interstellar turbulence. This scenario, initially proposed by \citet{Elitzur1978a,Elitzur1980}, was studied using 1D idealized structures of turbulent dissipation such as MHD shocks (e.g., \citealt{Draine1986, Pineau-des-Forets1986a, Flower1998c}) or magnetized vortices \citep{Godard2009,Godard2014}. In particular, \citet{Godard2014} found that the observed abundances of \CHp\ and many other molecular species can be explained if (i) all the mechanical energy of the CNM is dissipated in the CNM itself, (ii) in structures where the dissipation occurs through ambipolar diffusion with (iii) a typical ion-neutral velocity drift of $\sim 3$ \kms, three stringent necessary conditions. 

These studies were recently followed by 2D numerical simulations of hydrodynamic turbulence \citep{Lesaffre2020} and 3D numerical simulations of ideal MHD turbulence \citep{Myers2015,Moseley2021} applied to the CNM. \citet{Myers2015} and \citet{Moseley2021} argued that the distribution of the ion-neutral velocity drift expected in the CNM is sufficient to explain the observed abundances of \CHp. However their estimation of the velocity drift, obtained in the framework of ideal MHD, with no feedback on the magnetic field strength, constant ionization and molecular fractions, and a single momentum transfer rate coefficient, leads to artificially large velocity drifts in the low density gas ($\dens \sim 1$ \cc) where most of the \CHp\ is produced. As pointed out by \citet{Moseley2021}, the effect is so large that an arbitrary cut-off at 5 \kms\ needs to be applied to prevent the ambipolar heating rate to exceed the driving power of their simulations. Moreover, by focusing on the CNM, these simulations neglect the transfer of kinetic energy between the CNM and the WNM that naturally occurs in multiphase environments.

The above discussion shows that while the scenario of turbulent dissipation offers a plausible solution, it suffers from two major caveats: (1) it surmises that the mechanical energy dissipated in the CNM is of the order of the mechanical energy of CNM clouds, which is debatable ; (2) it requires reliable descriptions of the distribution of the ion-neutral velocity drift in regions of turbulent dissipation which are currently uncertain and overestimated in the framework of ideal MHD.

Another possible, yet poorly explored, solution to the formation of \CHp\ in the diffuse ISM is the exchange of matter between the WNM and the CNM induced by turbulent mixing \citep{Lesaffre2007} or a combination of turbulent advection and thermal instability \citep{Valdivia2017}. This scenario invokes the fact that cold molecular hydrogen initially formed in the CNM is naturally transported into the warm unstable phase where it survives long enough to activate the formation of \CHp. \citet{Valdivia2017} showed that this process greatly increases the production of \CHp\, although to a level 3 to 10 times lower than the observations. This last result was derived, however, from the outputs of a single simulation with no exploration of the parameter domain. In particular, the distribution of the column densities of \HH\ predicted by the simulation was in poor statistical agreement with the observations of the HI-to-\HH\ transition in the local ISM (see Fig.~13 of \citealt{Valdivia2016}). In addition, the comparisons with the observations were performed without taking into account the distribution of lengths of the observed diffuse lines of sight.

All these limitations were recently obviated by \citet{Bellomi2020} (hereafter \citetalias{Bellomi2020}) who performed a large parametric study, including 305 numerical simulations, of the turbulent and multiphase diffuse ISM. The results of their standard simulation were shown to simultaneously explain the position, the width, the dispersion, and most of the statistical properties of the HI-to-\HH\ transition observed in the local ISM with a precision never achieved by any previous theoretical model. The parametric study led to the conclusion that the observed HI-to-\HH\ transition is statistical in nature and traces the distributions and sizes of warm and cold phases mostly set by the mean density of the local diffuse ISM and the density of OB stars, with little dependence on the strength or the nature of the turbulent forcing. 

In the present paper, we reinvestigate the scenario of formation of \CHp\ induced by the turbulent and multiphase nature of the interstellar medium applying the methodology of \citetalias{Bellomi2020}. The observational sample is presented in Sect. \ref{Sect-Obs}. The setup, the updates, and the main properties of the numerical simulations are described in Sect. \ref{Sect-Sim}. The comparison with the observations and the exploration of the parameters are performed in Sects. \ref{Sect-Res} and \ref{Sect-params}. Discussion of our results and the main conclusions are presented in Sects. \ref{Sect-Dis} and \ref{Sect-Con}.

\section{Observational sample} \label{Sect-Obs}

\begin{figure}[!h]
\begin{center}
\includegraphics[width=9.0cm,trim = 2cm 1cm 0.5cm 3.0cm, clip,angle=0]{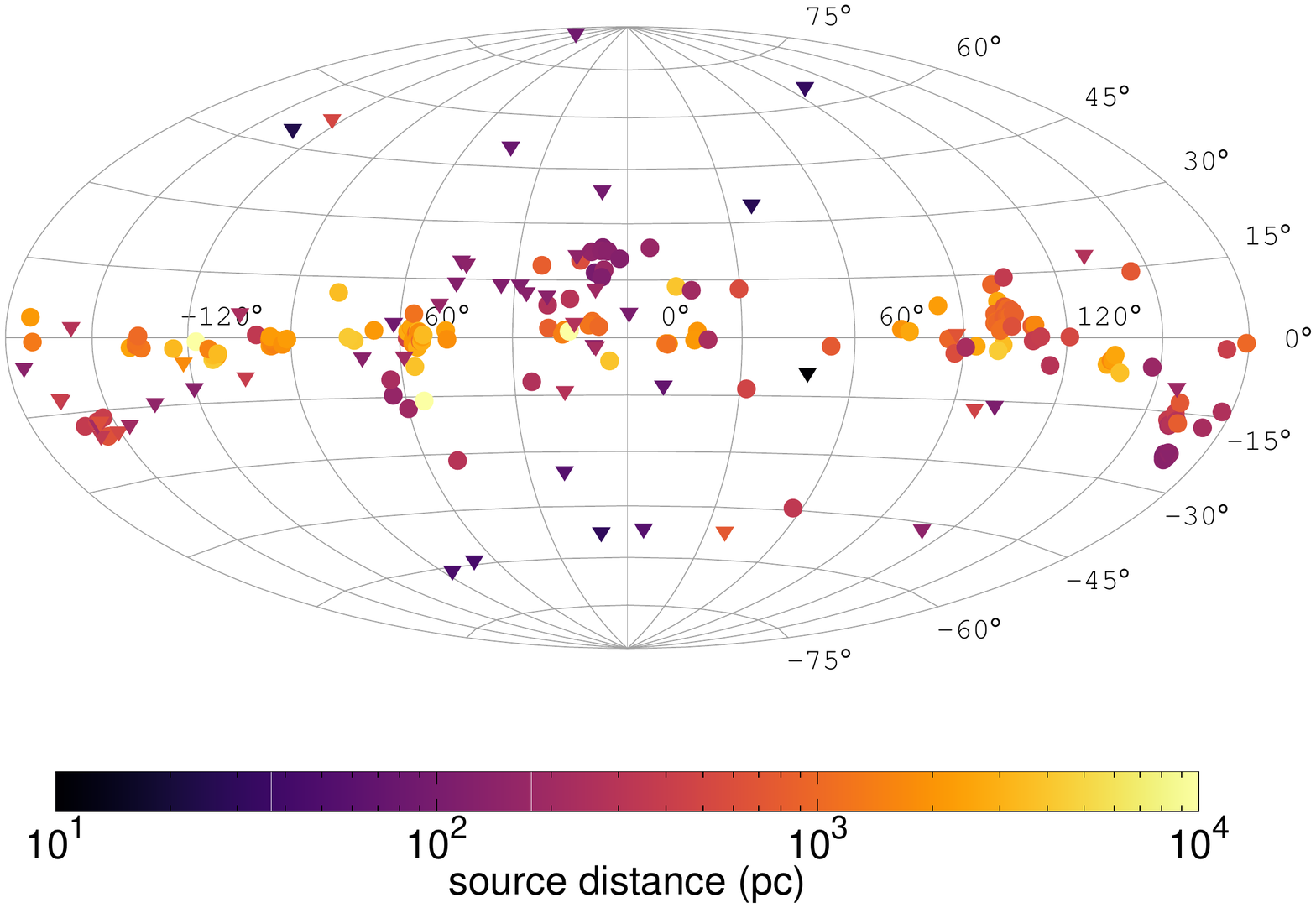}
\includegraphics[width=9.0cm,trim = 2cm 1cm 0.5cm 3.0cm, clip,angle=0]{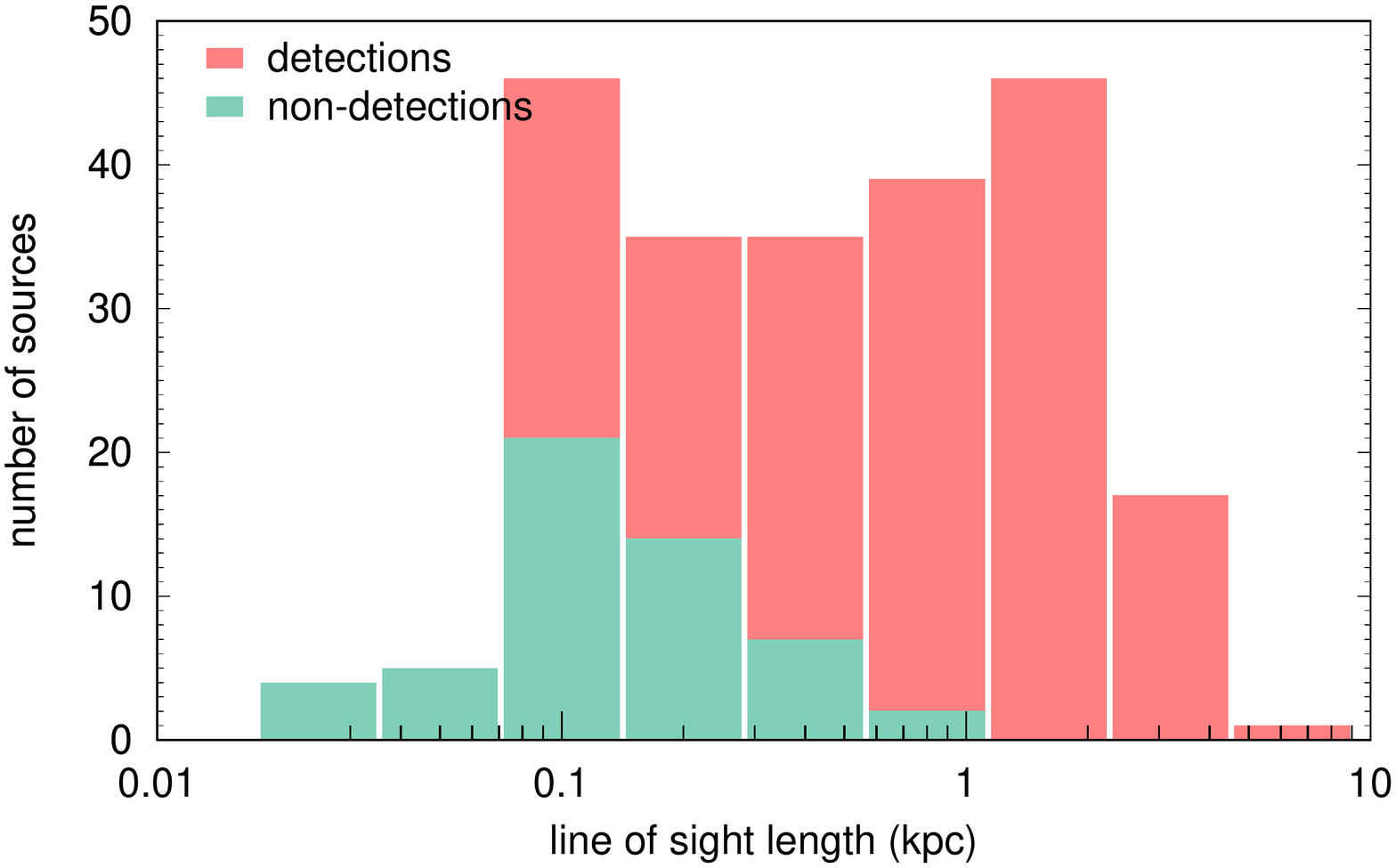}
\caption{{\it Top:} Aitoff projection, in Galactic longitude and latitude coordinates, of the background sources of the \CHp\ observational sample (see Table~\ref{Tab-Obs} of Appendix \ref{AppendObs}). The color code indicates the distances of background sources and the types of points the lines of sight where \CHp\ is detected (circles) or not detected (triangles). {\it Bottom:} Distribution of the lengths of the intercepted diffuse material computed with Eq. \ref{Eq-length} along all the lines of sight of the observational sample. The orange sample corresponds to lines of sight where \CHp\ is detected and the green sample to those for which an upper limit on $N(\CHp)$ has been derived (see the bottom part of Table~\ref{Tab-Obs}). The two samples are summed in order to display the total number of sources observed in each bin.}
\label{Fig-position-source}
\end{center}
\end{figure}

The observational sample is described in details in Appendix \ref{AppendObs} and Table~\ref{Tab-Obs}. It contains 229 lines of sight where the 4232 \AA\ line of \CHp\ was searched in absorption against the optical continuum of background stars. This sample, which forms to date the most complete set of observations of \CHp\ in the local diffuse ISM, includes 175 detections and 54 upper limits. The position of the sources in Galactic coordinates and their distances derived from recent measurements of their parallaxes are shown in the top panel of Fig.~\ref{Fig-position-source}. The lines of sight are rather homogeneously distributed in Galactic longitude. In contrast, and similarly to the sample studied in \citetalias{Bellomi2020}, about two-third of the sources are located at a Galactic latitude below 15$^\circ$ following the distribution of stars in the Solar Neighborhood. 

Because the amount of gas exponentially decreases as a function of the distance from the midplane of the Galaxy, the length of material intercepted by any line of sight is not always equal to the distance of the background source. Following the simple prescription adopted in \citetalias{Bellomi2020}, we assume a height above the midplane of 100 pc and compute this length as
\begin{equation} \label{Eq-length}
{\rm min}\left(d, \frac{100}{{\rm sin}(|b|)} \right)\,\, {\rm pc},
\end{equation}
where $d$ is the distance of the background source and $b$ is its Galactic latitude (see Table~\ref{Tab-Obs}). The distribution of lengths of the intercepted diffuse gas is shown in the bottom panel of Fig.~\ref{Fig-position-source} and extend over more than two orders of magnitude, from $\sim 10$ pc to a few kpc. We find that the length of the intercepted diffuse material follows a flat distribution in log scale from 0.1 to 2 kpc, as was found in for the observations of the HI-to-\HH\ transitions in the Solar Neighborhood \citepalias{Bellomi2020}. Interestingly, \CHp\ is never detected along lines of sight shorter than $\sim 100$ pc. Conversely, the number of non-detections of \CHp\ rapidly decreases with the length of the line of sight and reaches zero at $\sim 1$ kpc. As explained in \citetalias{Bellomi2020}, this distribution of length of the intercepted material is of paramount importance to accurately compare the results of numerical simulations with the distribution of chemical composition observed in the diffuse ISM.

\section{Theoretical models of the multiphase ISM} \label{Sect-Sim}

\subsection{Numerical setup}

\begin{table}
\caption{Standard values of the simulation parameters.}
\centering
\begin{tabular}{l c r l} 
\hline
\multicolumn{4}{c}{\vspace{-0.3cm}} \\
parameter              & symbol & value & Unit \\
\multicolumn{4}{c}{\vspace{-0.3cm}} \\
\hline
\multicolumn{4}{c}{\vspace{-0.3cm}} \\
box size               & $L$                 & 200                   & pc             \\
mean density           & $\overline{\dens}$  & 1.5                   & \cc            \\
UV radiation           & $G_0$               & 1                     & Mathis field   \\
resolution             & $R$                 & $512^3$               &                \\
forcing strength       & $F$                 & $1.5 \times 10^{-3}$  & kpc Myr$^{-2}$ \\
compressive ratio      & $\zeta$             & $0.1$                 &                \\
initial magnetic field & $B_x$               & $3.8$                 & $\mu$G         \\
\multicolumn{4}{c}{\vspace{-0.3cm}} \\
\hline
\end{tabular}
\label{tab-params}
\end{table}

The theoretical models used in this work are updates of the large grid of MHD  simulations of the multiphase diffuse ISM presented in length in \citetalias{Bellomi2020}. In a nutshell, the diffuse, turbulent, and multiphase neutral medium is modeled using the RAMSES code \citep{Teyssier2002,Fromang2006}. The simulations follow a magnetized and partially ionized gas with a mean proton density $\overline{\dens}$ over a box of size $L$ with periodic boundary conditions and illuminated from all sides by the standard isotropic UV radiation field \citep{Mathis1983} scaled by a factor $G_0$. The $z$ axis is chosen as the direction perpendicular to the Galactic midplane. A $z$-dependent gravitational potential is included to mimic the presence of stars and dark matter, while an homogeneous initial magnetic field $B_x$ is set along the $x$ axis. Throughout the simulations, mechanical energy is regularly injected in the gas at large scale ($\sim L/2$) via random fluctuations of the acceleration field in Fourier space (e.g., \citealt{Schmidt2009,Federrath2010}). The amplitude of this forcing and the relative power injected in compressive modes are controlled by two parameters, $F$ and $\zeta$ respectively (e.g., \citealt{Saury2014}). All the models are run without adaptative mesh refinement: the number of grid cells is therefore equal to the numerical resolution $R$.

In this configuration, RAMSES follows the thermodynamical evolution of the gas, taking into account heating and cooling processes, and the out-of-equilibrium evolution of \Hp, H, and H$_2$ as described below. Each simulation is run over a few turnover timescales until it reaches a steady-state where the physical properties of the multiphase ISM (the fraction of gas in the different phases, the mean pressure, the mean velocity dispersion, and the mean molecular fraction) are roughly stationary. The standard set of parameters is given in Table~\ref{tab-params}. It slightly differs from that adopted in \citetalias{Bellomi2020}, with a mean density  $\overline{\dens} = 1.5$ \cc\ (instead of 2 \cc), a turbulent forcing strength $F = 1.5 \times 10^{-3}$ kpc Myr$^{-2}$ (instead of $9 \times 10^{-4}$ kpc Myr$^{-2}$) and a resolution of $512^3$ (instead of $256^3$).

Several physical and chemical processes have been modified compared to those treated in the simulations used in \citetalias{Bellomi2020}. These modifications, described in the two following sections, include updates of the heating and cooling processes, the calculation of the time dependent evolution of the abundance of \Hp, and updates of the treatment of the photodissociation of \HH.

\subsection{Updates of the heating and cooling processes}

\begin{table*}[!h]
\caption{Reaction rates of the formation and destruction processes of \Hp, PAH$^-$, PAH$^0$, and PAH$^+$ included in the simulations.}
\centering
\begin{tabular}{l @{\hspace{0.1cm}} 
                l @{\hspace{0.1cm}}
                l @{\hspace{0.1cm}}
                l @{\hspace{0.1cm}}
                l @{\hspace{0.1cm}}
                l @{\hspace{0.1cm}}
                l l l l} 
\hline
\multicolumn{9}{c}{\vspace{-0.3cm}} \\
\multicolumn{7}{c}{reaction} & rate & unit & reference \\
\multicolumn{9}{c}{\vspace{-0.3cm}} \\
\hline
\multicolumn{9}{c}{\vspace{-0.3cm}} \\
H       & + & CR       & $\rightarrow$ & \Hp     &   &         & $2.0 \times 10^{-16}$                            & s$^{-1}$        & \citet{Wolfire2003} total ionization rate$^{(a)}$ \\
\Hp     & + & PAH$^-$  & $\rightarrow$ & H       & + & PAH$^0$ & $8.3 \times 10^{-07}\,\,(T/100 {\rm K})^{-0.5}$  & cm$^3$ s$^{-1}$ & \citet{Draine1987}$^{(b)}$                        \\
\Hp     & + & PAH$^0$  & $\rightarrow$ & H       & + & PAH$^+$ & $3.1 \times 10^{-08}$                            & cm$^3$ s$^{-1}$ & \citet{Draine1987}$^{(b)}$                        \\
PAH$^-$ & + & $\gamma$ & $\rightarrow$ & PAH$^0$ & + & e$^-$   & $1.8 \times 10^{-08}\,\,{\rm exp}(-2.5 A_V)$     & s$^{-1}$        & \citet{Weingartner2001a}$^{(c)}$                   \\
PAH$^0$ & + & $\gamma$ & $\rightarrow$ & PAH$^+$ & + & e$^-$   & $1.2 \times 10^{-08}\,\,{\rm exp}(-2.5 A_V)$     & s$^{-1}$        & \citet{Weingartner2001a}$^{(c)}$                   \\
PAH$^0$ & + & e$^-$    & $\rightarrow$ & PAH$^-$ &   &         & $1.3 \times 10^{-06}$                            & cm$^3$ s$^{-1}$ & \citet{Draine1987}$^{(b,d)}$                   \\
PAH$^+$ & + & e$^-$    & $\rightarrow$ & PAH$^0$ &   &         & $3.5 \times 10^{-05}\,\,(T/100 {\rm K})^{-0.5}$  & cm$^3$ s$^{-1}$ & \citet{Draine1987}$^{(b,d)}$                   \\
\multicolumn{10}{c}{\vspace{-0.3cm}} \\
\hline
\multicolumn{10}{c}{\vspace{ 0.0cm}} \\
\multicolumn{10}{p{15cm}}{
{\bf Notes.} $^{(a)}$ This rate includes the ionization induced by EUV photons, X-rays, and cosmic rays. $^{(b)}$ Rates computed for a spherical PAH with a radius of 5 \AA. $^{(c)}$ The dependence on the visual extinction $A_V$ is set to match the dependence on $A_V$ of the photoelectric effect adopted in Eq. 11 of \citetalias{Bellomi2020}. $^{(d)}$ Assuming a sticking coefficient of electrons onto PAHs, $s_e=1$.}
\end{tabular}
\label{tab-Hp-PAH}
\end{table*}

In \citetalias{Bellomi2020}, the thermodynamical evolution of the gas was computed taking into account the heating induced by the photoelectric effect and the cosmic ray particles, the radiative cooling induced by the Lyman $\alpha$ line and the fine structure lines of OI and CII, and the cooling due to the recombination of electrons onto grains. 

To account for the impact of \HH\ on the thermodynamical evolution of the gas, we include here three additional mechanisms, the heating induced by the formation of \HH, the heating that follows its photodestruction, and the radiative cooling induced by the collisional excitation of its rovibrational levels. The first two heating rates are modeled using Eqs. 13 and 14 of \citet{Valdivia2016} who assumed that a third of the binding energy of \HH\ (4.5 eV) is released in the gas during its formation and that 0.4 eV is released in the gas per photodestruction \citep{Black1977}. The cooling induced by the collisional excitation of \HH\ is calculated using the analytical functions prescribed by \citet{Moseley2021} (Eqs. 17, 18, 19, 20, and 21 of their paper) who performed fits of the cooling rate of \HH, calculated with updated collisional rate coefficients, valid for kinetic temperatures below 5000 K.

Regarding the photoelectric effect, we adopt here the photoelectric heating rate given by \citet{Weingartner2001a} who provide a specific prescription for the diffuse interstellar medium (see Eq. 44 of their paper and the last line of their Table 2). Following \citet{Wolfire2003}, and to insure self-consistency between the computations of the photoelectric heating and the fractional abundance of \Hp\ (see Sect. \ref{Sect-chem}), we adopt a grain charging parameter
\begin{equation} \label{Eq-charg-param}
\Upsilon = \frac{1.3\,\,G_0 \sqrt{T}}{n_{\rm e} s_{\rm e} / 0.164},
\end{equation}
that depends on the sticking coefficient of electrons onto PAHs, $s_{\rm e}$. In the above formula, $T$ is the kinetic temperature of the gas, and $n_{\rm e}$ is the density of electrons. 
The resulting photoelectric heating rate therefore writes
\begin{equation} \label{Eq-photo-heat}
10^{-26} \frac{1.3\,\,G_0 \,\, \dens \,\, \left( 5.45 + 2.5\,\,T^{0.147} \right)}{1 + 0.00945\,\, \Upsilon^{0.623} \left(1 + 0.01453\,\,\Upsilon^{0.511} \right)} \,\, {\rm erg}\,\,{\rm cm}^{-3}\,\,{\rm s}^{-1}.
\end{equation}
The factor 1.3 in Eqs. \ref{Eq-charg-param} and \ref{Eq-photo-heat} accounts for the ratio between the energy density of the Mathis field \citep{Mathis1983}, used as reference in this work, and that of the Habing field \citep{Habing1968}, used as reference by \citet{Weingartner2001a}. Throughout this work, the sticking coefficient of electrons onto PAHs is set to 1 for the computations of both the photoelectric heating rate and the out-of-equilibrium abundance of \Hp.


\subsection{Treatment of the chemistry} \label{Sect-chem}

The timescale required for the electron abundance to reach chemical equilibrium spans several orders of magnitude depending on the physical conditions of the ISM. Expected to be short for CNM conditions ($\sim 10^4$ yr), this equilibrium timescale considerably increases in WNM environments (a few $10^6$ yr) where it becomes larger than the dynamical timescale. To account for potential out-of-equilibrium effects in the unstable and the WNM phases, we model the fractional abundance of electrons as $x({\rm e}^-) = x(\Cp) + x(\Hp)$, where the fractional abundance of \Cp, $x(\Cp)$, is set to a constant value of $1.4 \times 10^{-4}$, while that of \Hp, $x(\Hp)$, is computed in the simulation. Following \citet{Wolfire2003}, we assume that the fractional abundance of H$^+$ in the diffuse neutral ISM is driven by the ionization of H induced by EUV photons, soft X-rays, and cosmic rays, and by the recombination of H$^+$ on negatively charged and neutral PAHs\footnote{In particular, the radiative recombination of H$^+$ and the charge exchange of H$^+$ with neutral Oxygen are neglected.}. The abundances of PAHs in their different ionization states are supposed to be driven by photodetachment and photoionization processes, on the one side, and by recombination with free electrons, on the other side. The rates adopted for all these processes are given in Table \ref{tab-Hp-PAH}. The size of the PAHs is set to 5 \AA\ and their fractional abundance to $10^{-6}$ which corresponds to a total abundance of Carbon of $6 \times 10^{-5}$ in very small grains \citep{Weingartner2001a}. In diffuse interstellar conditions, the timescale required for the PAHs, in any ionized state, to reach their equilibrium abundance is smaller than 10 yr which is far shorter than the dynamical timescale of the gas and the equilibrium timescale of \Hp. The abundances of negatively charged, neutral, and positively charged PAHs can therefore be computed at equilibrium. This consideration leads to a simplified differential equation for the abundance of \Hp\ which is solved in each cell and at each timestep in the simulation using the splitting operator method.

As in \citetalias{Bellomi2020}, the out-of-equilibrium evolution of the abundances of H and \HH\ are computed in the simulation, taking into account the formation of \HH\ onto grains and its photodestruction by UV photons. In particular, the photodestruction rate of \HH\ is modeled as
\begin{equation} \label{Eq-H2des}
3.3 \times 10^{-11} \,\, G_0 \,\, f_{\rm sh, \HH}  \,\, {\rm s}^{-1},
\end{equation}
where the shielding factor, $f_{\rm sh, \HH},$ is computed by averaging at any point the shielding factor due to dust extinction and the self-shielding factor of \HH\ over 12 solid angles evenly spread in polar coordinates \citep{Valdivia2016}. In \citetalias{Bellomi2020}, the self-shielding of \HH\ was derived using the prescription of \citet{Draine1996}. Such a prescription is reliable for diffuse gas at low temperature but becomes less and less reliable for high temperature environments ($T > 500$ K) where the collisional excitation of \HH\ in its rovibrational levels reduces the efficiency of the self-shielding \citep{Wolcott-Green2011}. To account for this effect, we use here the self-shielding function given by \citet{Wolcott-Green2011} (Eq. 12 in their paper with $\alpha=1.1$) using a Doppler broadening parameter of 2 \kms\ in favor of the small-scale self-shielding (see Sects. 5.5 and 5.6 of \citetalias{Bellomi2020}).

\begin{table*}[!h]
\caption{Reaction rates of the main formation and destruction processes of \CHp\ and associated references.}
\centering
\begin{tabular}{l @{\hspace{0.1cm}} 
                l @{\hspace{0.1cm}}
                l @{\hspace{0.1cm}}
                l @{\hspace{0.1cm}}
                l @{\hspace{0.1cm}}
                l @{\hspace{0.1cm}}
                l l l l} 
\hline
\multicolumn{9}{c}{\vspace{-0.3cm}} \\
\multicolumn{7}{c}{reaction} & rate & unit & ref \\
\multicolumn{9}{c}{\vspace{-0.3cm}} \\
\hline
\multicolumn{9}{c}{\vspace{-0.3cm}} \\
\Cp  & + & \HH      & $\rightarrow$ & \CHp  & + & H   & $7.4 \times 10^{-10}\,\,{\rm exp}(-4537/kT)$   & cm$^3$ s$^{-1}$ & \citet{Hierl1997}  \\
\CHp & + & H        & $\rightarrow$ & \Cp   & + & \HH & $7.8 \times 10^{-10} (T/300 {\rm K})^{-0.22}$  & cm$^3$ s$^{-1}$ & \citet{Plasil2011} \\
\CHp & + & \HH      & $\rightarrow$ & \CHdp & + & H   & $1.2 \times 10^{-09}$                          & cm$^3$ s$^{-1}$ & \citet{McEwan1999} \\
\CHp & + & $\gamma$ & $\rightarrow$ & C     & + & \Hp & $2.5 \times 10^{-10}\,\,{\rm exp}(-3.5 A_V)$   & s$^{-1}$        & \citet{Heays2017} \\
\multirow{5}{*}{\CHp} & \multirow{5}{*}{+} & \multirow{5}{*}{e$^-$} & \multirow{5}{*}{$\rightarrow$} & \multirow{5}{*}{C} & \multirow{5}{*}{+} & \multirow{5}{*}{H} & $2.43 \times 10^{-07} (T/300 {\rm K})^{-0.74} + $           & \multirow{5}{*}{cm$^3$ s$^{-1}$} & \multirow{5}{*}{\citet{Paul2022}} \\
                      &                    &                        &                                &                    &                    &                    & $6.42 \times 10^{-01} T^{-1.5} {\rm exp}(-112000/T)\,\, + $ &                                  &                                                \\
                      &                    &                        &                                &                    &                    &                    & $2.36 \times 10^{-02} T^{-1.5} {\rm exp}(-12000/T)\,\,- $   &                                  &                                                \\
                      &                    &                        &                                &                    &                    &                    & $2.58 \times 10^{-03} T^{-1.5} {\rm exp}(-941/T)\,\, - $    &                                  &                                                \\
                      &                    &                        &                                &                    &                    &                    & $1.13 \times 10^{-03} T^{-1.5} {\rm exp}(-220/T)$           &                                  &                                                \\
\multicolumn{9}{c}{\vspace{-0.3cm}} \\
\hline
\end{tabular}
\label{tab-reactions}
\end{table*}

\begin{figure}[!h]
\begin{center}
\includegraphics[width=9.8cm,trim = 1.0cm 2.5cm 0.5cm 3.5cm, clip,angle=0]{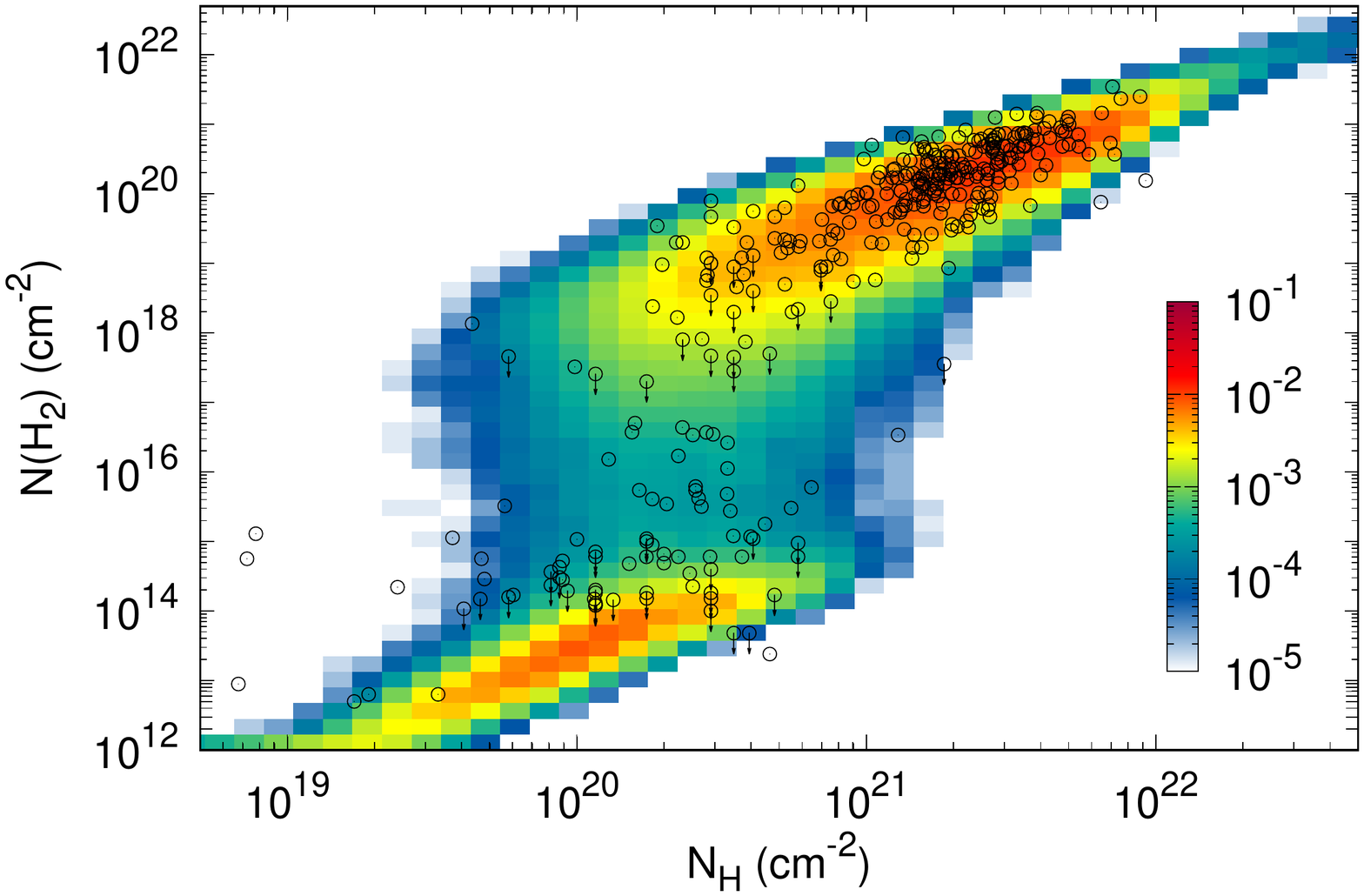}
\includegraphics[width=9.0cm,trim = 0.5cm 2.5cm 0.5cm 2.0cm, clip,angle=0]{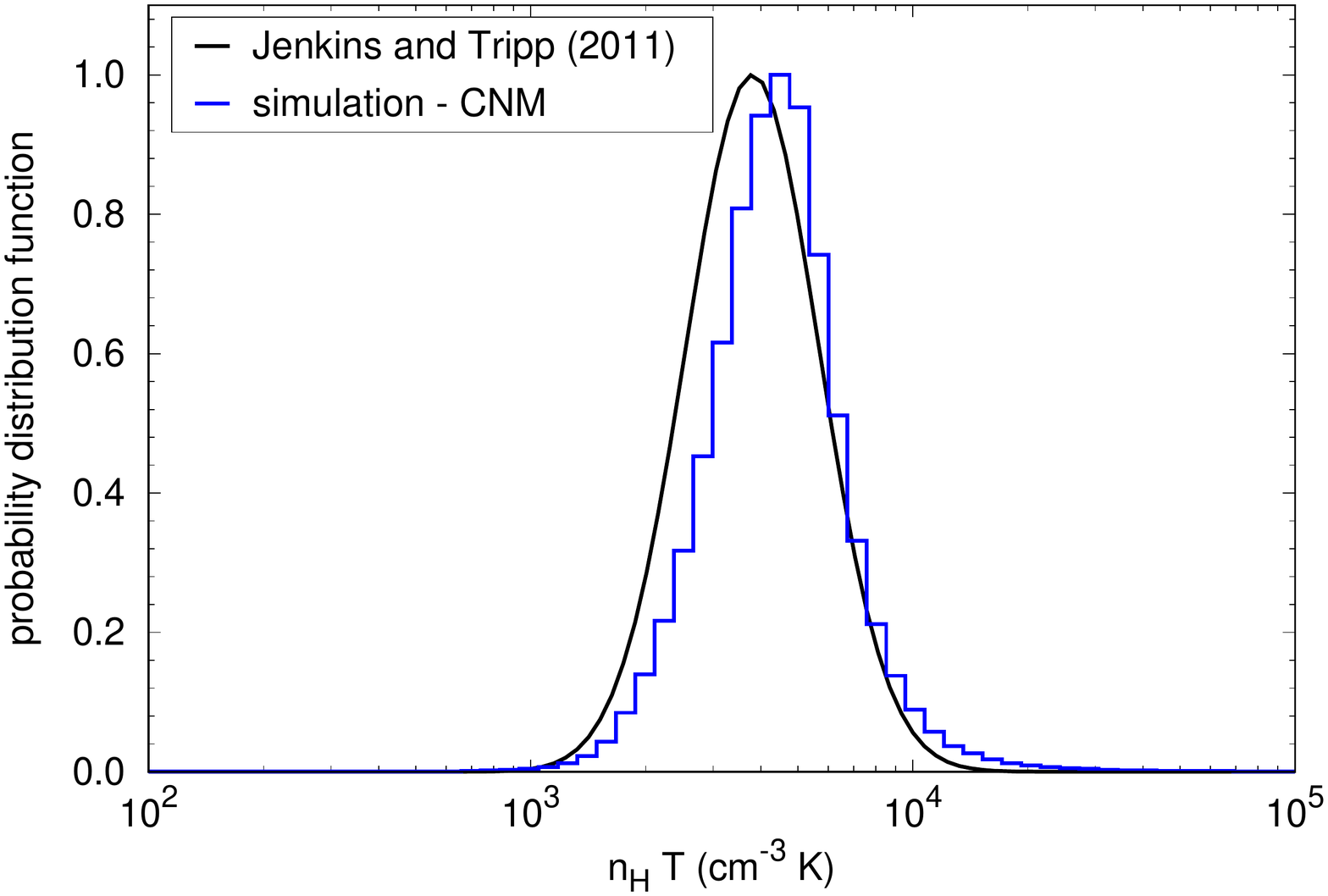}
\caption{{\it top}~: Comparison of the observational dataset of the HI-to-\HH\ transition of \citetalias{Bellomi2020} (black points) to the 2D probability histogram obtained with the standard simulation after application of the reconstruction algorithm (see Sect. 3.8 of \citetalias{Bellomi2020}). Observations include detections of \HH\ (circles) and upper limits on its column density (arrows). The color code indicates the fraction of lines of sight (in logarithmic scale) contained in each bin. {\it Bottom}~: Probability distribution function of the thermal pressure of the cold neutral medium in the standard simulation (blue histograms) compared with the probability distribution function of the thermal pressure deduced from the observation of the fine structure levels of Carbon \citet{Jenkins2011} (black curve). Both distributions are normalized to their maximum value to facilitate the comparison.}
\label{Fig-main-prop}
\end{center}
\end{figure}

Only the abundances of \Hp, H, and \HH\ are computed during the numerical simulation. The rest of the chemical composition of the diffuse and multiphase ISM is computed in post-treatment using a chemical solver presented in Appendix A of \citet{Valdivia2017}. This solver computes the abundances of atomic and molecular species, assuming chemical equilibrium for all species except for \Hp, H, and \HH\ which, depending on the choice of the user, can be given as input parameters or computed at equilibrium. The chemical network used by the solver is taken from the Meudon PDR code\footnote{version 1.5.2 available on \url{http://ism.obspm.fr}} \citep{Le-Petit2006} and includes 151 species interacting through a network of 2715 reactions. In this network, the chemistry of \CHp\ is mainly driven by five different reactions, its formation through the hydrogenation of \Cp\,
and its destruction by photodissociation and reactive collisions with H, \HH, and e$^-$. The rates adopted for these processes are given in Table~\ref{tab-reactions}.

\subsection{Physical properties of the standard simulation} \label{Sect-standard}

The updates described in the previous sections have an important impact on the thermochemical properties of the multiphase ISM simulated with RAMSES compared with the simulations presented in \citetalias{Bellomi2020}. In particular, the cooling induced by the excitation of the rovibrational levels of \HH\ suppresses the thermal instability process in fully molecular media. This cooling mechanism therefore efficiently prevents CNM clouds to evaporate back into the WNM as long as \HH\ is not, at least partly, destroyed. The updates performed on the photoelectric effect and the computation of the out-of-equilibrium abundance of \Hp\ are found to boost the efficiency of the photoelectric heating and produce a multiphase gas at higher thermal pressure than those reported in Fig. 4 of \citetalias{Bellomi2020}.

Given the differences between our two studies, we discuss here the choice of the standard setup and compare its main physical properties to those derived in the local diffuse ISM.
The standard setup corresponds to a neutral diffuse ISM extending over 200 pc, illuminated by the interstellar radiation field of \citet{Mathis1983} and with a mean density $\overline{\dens} = 1.5$ \cc (see Table \ref{tab-params}). As discussed in \citetalias{Bellomi2020} (Sect. 3.5), the scale of illumination of the gas is comparable to the typical distances between OB star associations in the Solar Neighborhood (e.g., \citealt{Zari2018}). Given the uncertainties on the volume filling factor of the fully ionized gas (see Sect. 5.3 of \citetalias{Bellomi2020}), the mean density is in agreement with the standard Galactic midplane density of HI at a galactocentric distance of 8.5 kpc \citep{Kalberla2009} derived from the measurement of the mass surface density of HI in the Solar Neighborhood (e.g. \citealt{Nakanishi2016}).

The standard simulation reaches steady-state in $\sim 30$ Myr. The WNM (defined here by $T>3000$ K), the CNM (defined here by $T<300$ K), and the unstable gas at intermediate temperature, the Lukewarm Neutral Medium (LNM), encompass respectively 35\%, 54\%, and 11\% of the total mass of the gas. We display in Fig. \ref{Fig-main-prop} the 2D probability histogram of the column densities of H and \HH\ obtained with the standard simulation\footnote{The 2D histogram is obtained by applying the reconstruction algorithm used in \citetalias{Bellomi2020} to account for the distribution of length of the intercepted diffuse material in the observational sample of H and \HH\ (Table A.1 of \citetalias{Bellomi2020}).} and the probability distribution function of the pressure (normalized to its maximum) predicted in the CNM. As found in \citetalias{Bellomi2020}, the multiphase ISM simulated with the RAMSES code reproduces, to an outstanding level, the statistical properties of the HI-to-\HH\ transition observed in the local interstellar medium, which include the position of the transition, its width, the dispersion of observations, and the occurrence of most of the lines of sight. In addition, and as shown in the bottom panel of Fig. \ref{Fig-main-prop}, the distribution of thermal pressure obtained in the cold neutral medium is in agreement with the pressure distribution derived by \citet{Jenkins2011} (Eq. 3 of their paper) from the observations of the excitation of the fine structure levels of atomic Carbon. This last result significantly differs from \citetalias{Bellomi2020} where the standard simulation was underpredicting the mean thermal pressure of the ISM by a factor $\sim 3.5$ (see Fig. 4 of \citetalias{Bellomi2020}). The width of the pressure probability distribution function is controlled by the strength of the turbulent forcing which is set here to $F = 1.5 \times 10^{-3}$ kpc Myr$^{-2}$. We find that such a turbulent forcing not only insures that the CNM covers the same thermal pressure range than that inferred from the observations but also leads to a 1D turbulent velocity dispersion of the WNM (calculated using Eqs. 17, 19, and 20 of \citetalias{Bellomi2020}) of $\sim 5.4$ \kms, in agreement with the velocity dispersion deduced from HI emission spectra at high Galactic latitude \citep{Kalberla2005,Haud2007}.

The turbulent forcing and the homogeneous magnetic field applied in the simulation lead to an equipartition between the kinetic and magnetic energy densities at a level of $\sim 1.6 \times 10^{-12}$ \ecc. In comparison, the impinging UV radiation field\footnote{The total interstellar radiation field, excluding the Cosmic Microwave Background, integrated over all wavelength has an energy density of $\sim 1.6 \times 10^{-12}$ erg~\cc\ (\citealt{Draine2011}, Table 12.1), in equipartition with the kinetic and magnetic energy densities of the standard simulation.} \citep{Mathis1983} 
has an energy density of $6.1 \times 10^{-14}$ erg~\cc\ between 6 and 13.6 eV \citep{Weingartner2001a}. These reservoirs of energy have, however, very different cycling timescales. The dissipation of the mechanical energy is almost entirely mediated by numerical viscosity. The cycling of the mechanical energy typically occurs over a turnover timescale, implying an artificial turbulent heating rate of $\sim 1.5 \times 10^{-27}$ erg \cc\ s$^{-1}$. Because of the fast cycling of the UV radiative energy, this turbulent heating rate is far smaller than the heating induced by the photoelectric effect $\sim 10^{-25}$ erg \cc\ s$^{-1}$ which effectively dominates the heating of the gas and sets its thermal energy density at a level of $9.9 \times 10^{-13}$ erg \cc. The main physical properties of the standard simulation are summarized in Table \ref{tab-main-prop}.

\begin{table}
\caption{Main physical properties of the standard simulation.}
\centering
\begin{tabular}{l c r} 
\hline
\multicolumn{3}{c}{\vspace{-0.3cm}} \\
quantities             & value & unit \\
\multicolumn{3}{c}{\vspace{-0.3cm}} \\
\hline
\multicolumn{3}{c}{\vspace{-0.3cm}} \\
WNM mass fraction          & $35 \%$               &       \\
LNM mass fraction          & $11 \%$               &       \\
CNM mass fraction          & $54 \%$               &       \\
WNM 1D velocity dispersion & $5.4$                 & \kms  \\
kinetic energy density     & $1.7 \times 10^{-12}$ & \ecc  \\
magnetic energy density    & $1.6 \times 10^{-12}$ & \ecc  \\
thermal energy density     & $9.9 \times 10^{-13}$ & \ecc  \\
UV energy density          & $6.1 \times 10^{-14}$ & \ecc  \\
radiative energy density   & $1.6 \times 10^{-12}$ & \ecc  \\
\multicolumn{3}{c}{\vspace{-0.3cm}} \\
\hline
\end{tabular}
\label{tab-main-prop}
\end{table}

\subsection{Origin of \CHp} \label{Sect-origin}



\begin{figure}[!b]
\begin{center}
\includegraphics[width=9.0cm,trim = 0.7cm 2.5cm 0.5cm 3.5cm, clip,angle=0]{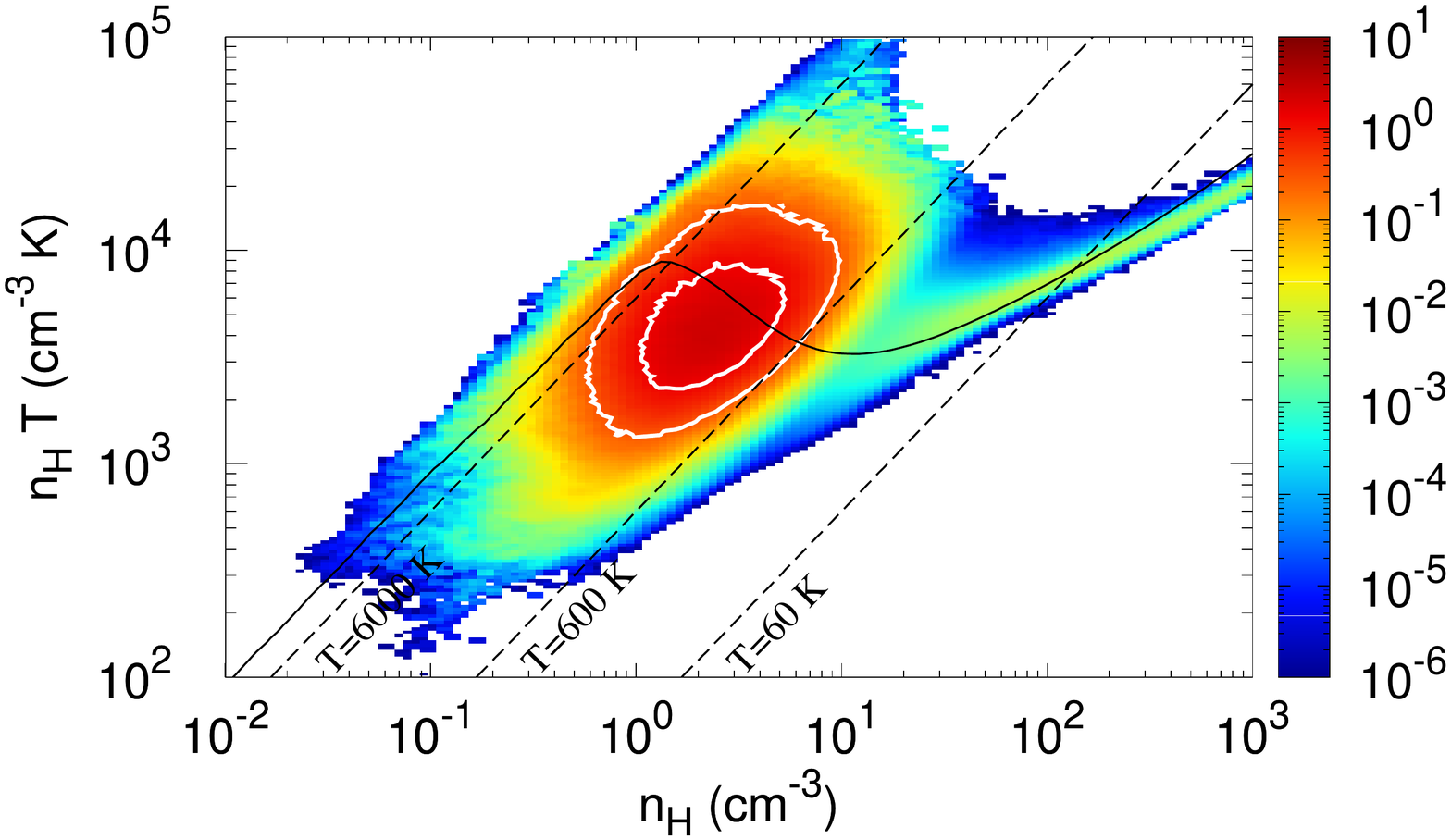}
\includegraphics[width=9.0cm,trim = 0.7cm 2.5cm 0.5cm 3.5cm, clip,angle=0]{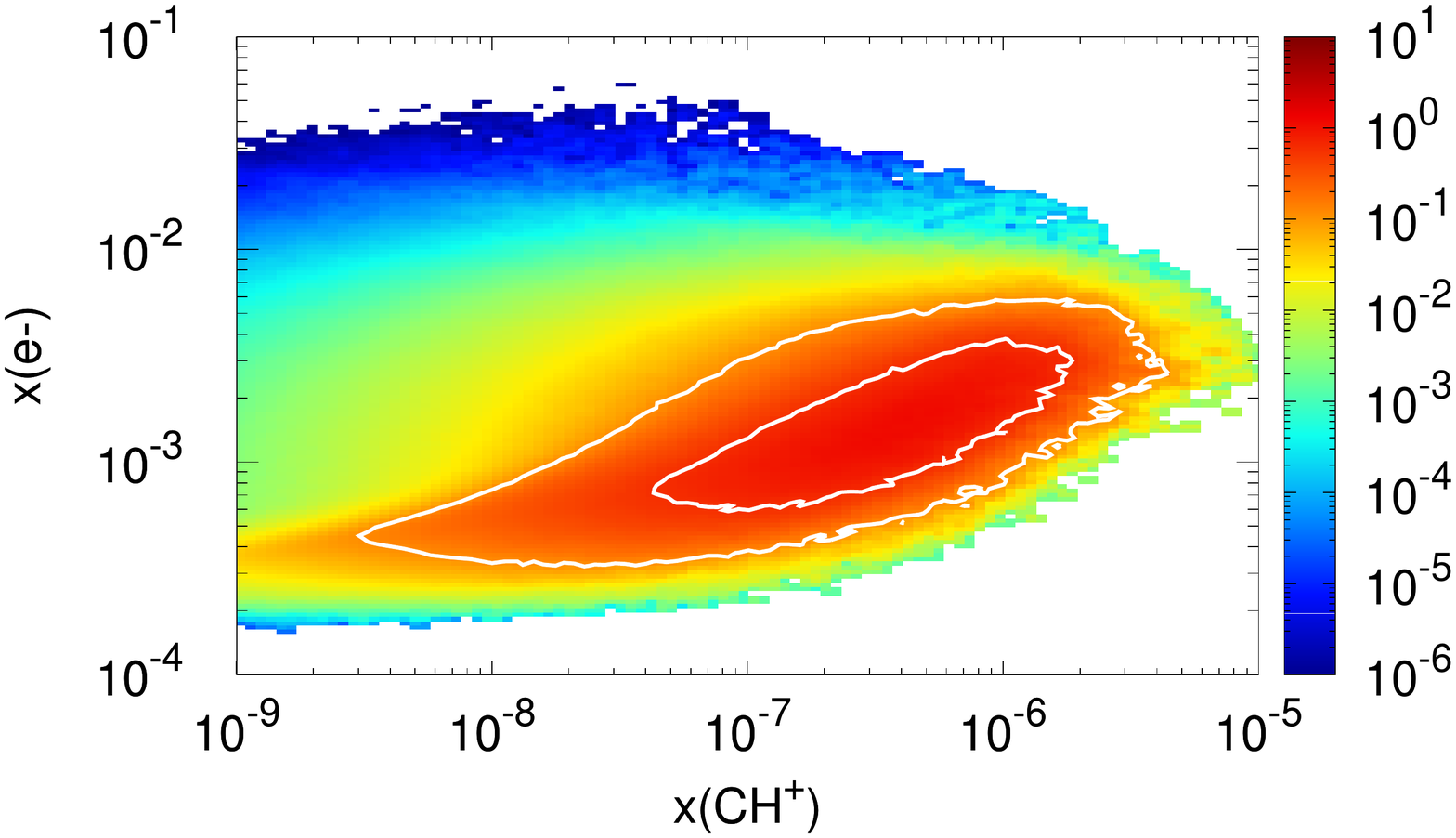}
\includegraphics[width=9.0cm,trim = 0.7cm 2.5cm 0.5cm 3.5cm, clip,angle=0]{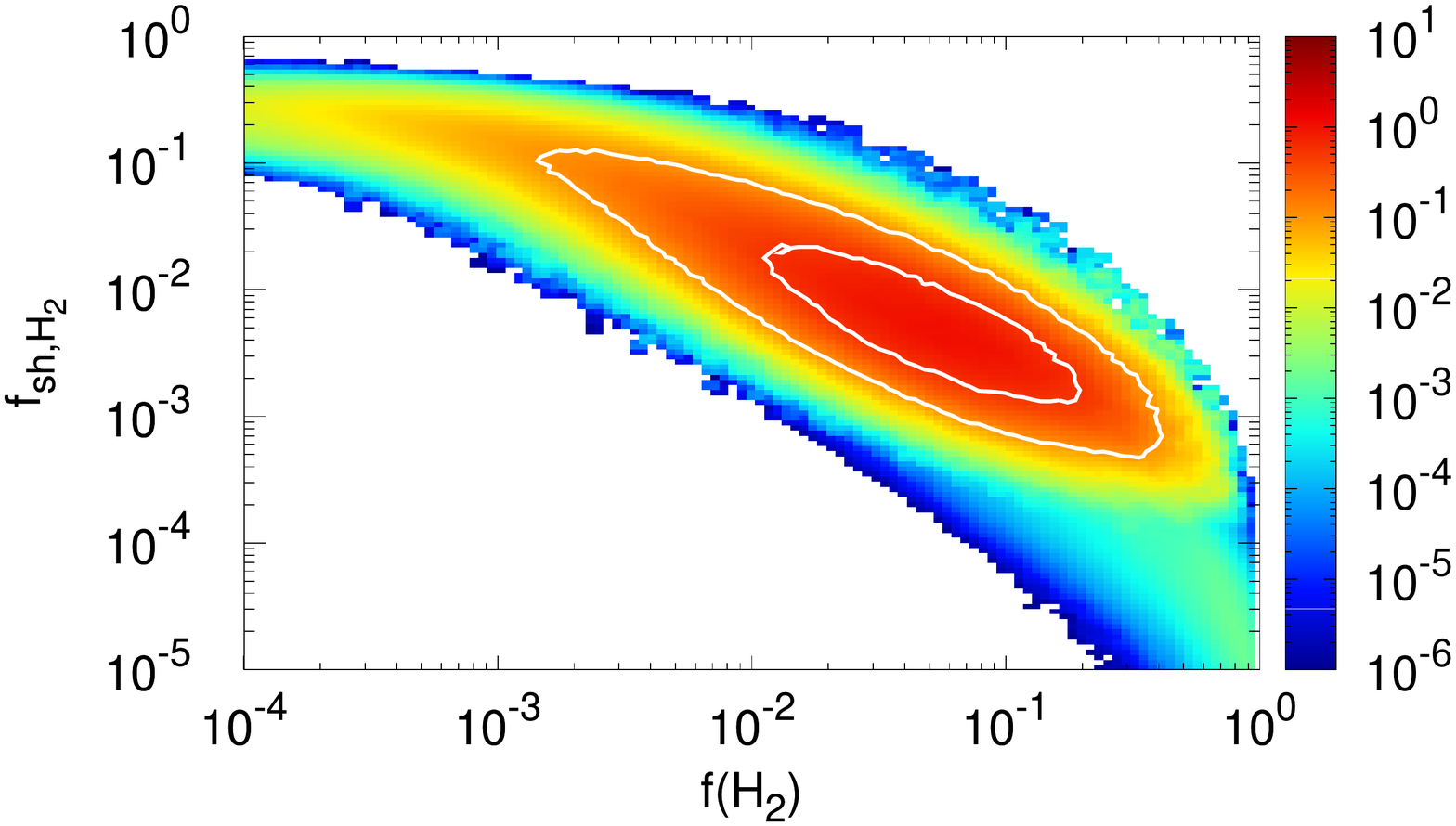}
\caption{2D probability distribution functions of the fraction of mass of \CHp\ in the fiducial simulation $d [M(\CHp)/M_{\rm tot}(\CHp)]$ / $[d{\rm log}(a)\,d{\rm log}(b)]$ displayed in a $a=\dens$ and $b=\dens T$ diagram (top panel), a $a=x(\CHp)$ and $b=x({\rm e}^-)$ diagram (middle panel), and a $a=f(\HH)$ and $b=f_{\rm sh, \HH}$ diagram (bottom panel). The black solid curve in the top panel indicates the thermal equilibrium state expected for a gas illuminated by the Mathis interstellar radiation field and located at a visual extinction of 0.2 magnitude which roughly corresponds to the visual extinction at the center of the standard simulation. The black dashed lines are isothermal contours at $T=60$, $600$, and $6000$ K. The white contours in all panels show the isocontours that encompass 50\% and 90\% of the total mass of \CHp, $M_{\rm tot}(\CHp)$.}
\label{Fig-form-dist}
\end{center}
\end{figure}

\begin{figure*}[!h]
\begin{center}
\includegraphics[width=9.0cm,trim = 0.0cm 5.0cm 1.0cm 9.0cm, clip,angle=0]{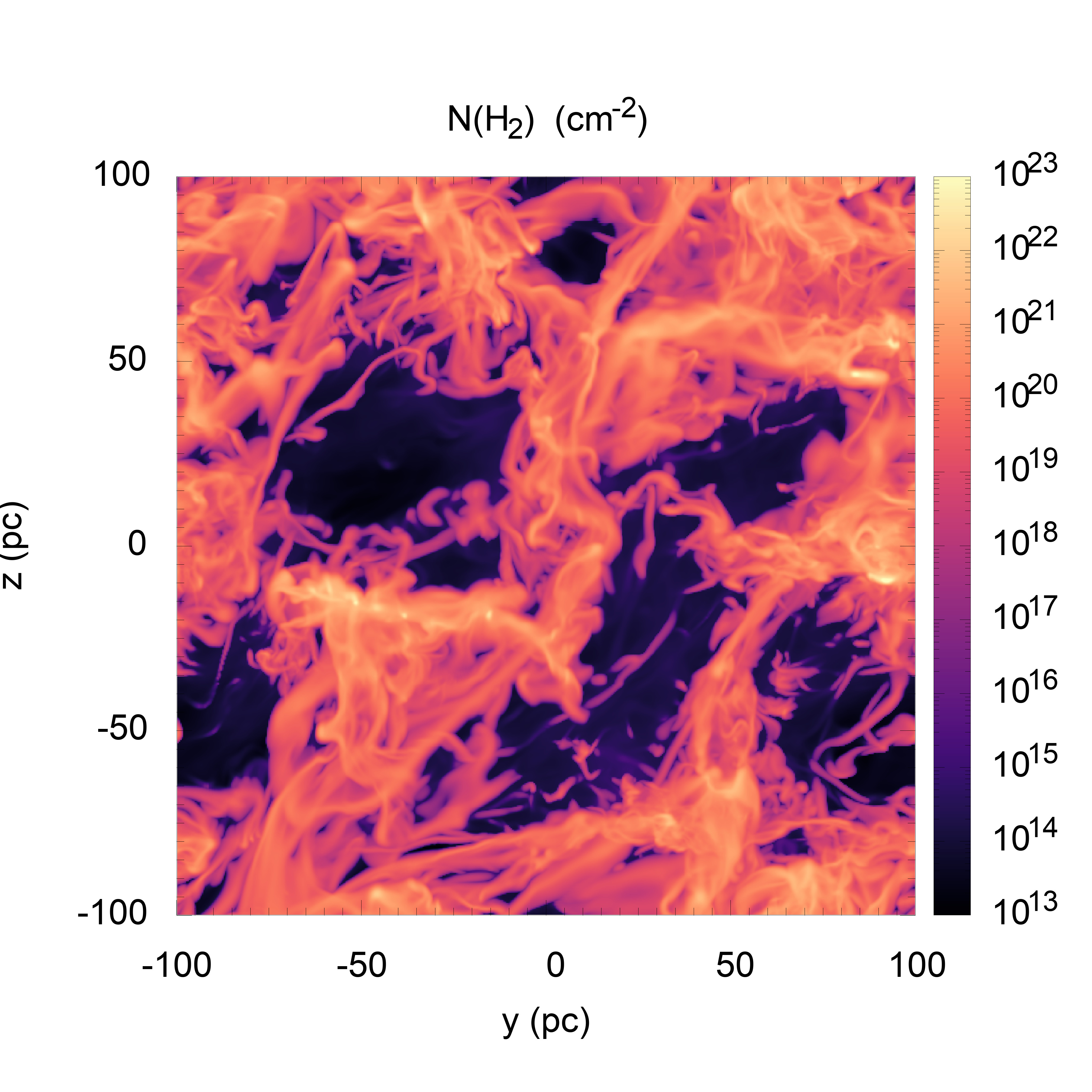}
\includegraphics[width=9.0cm,trim = 0.0cm 5.0cm 1.0cm 9.0cm, clip,angle=0]{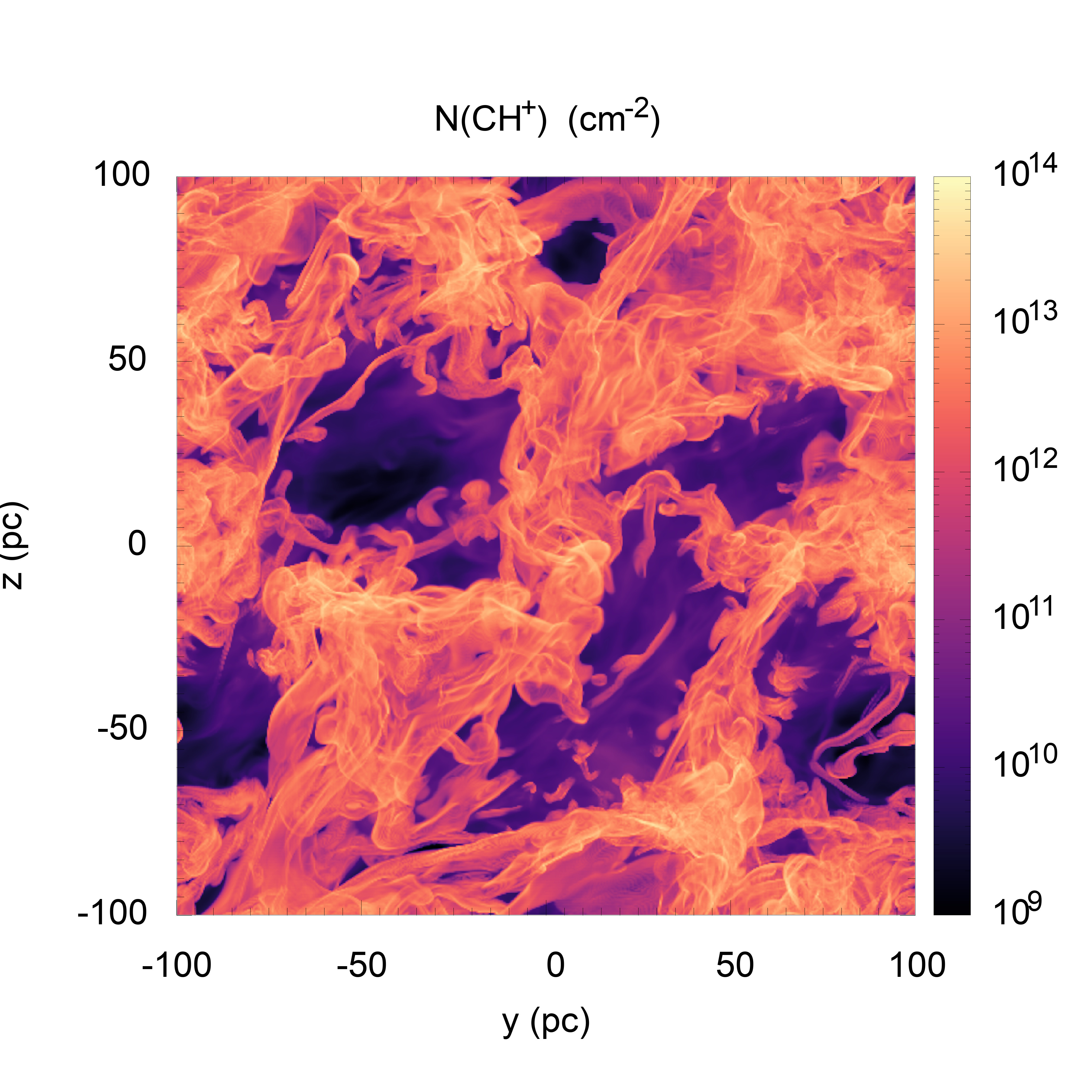}
\caption{Maps of the column densities of \HH\ (left panel) and \CHp\ (right panel) predicted with the standard simulation (see Table~\ref{tab-params}) and obtained by integrating the densities along the $x$ axis over the entire box of size $L=200$ pc.}
\label{Fig-map}
\end{center}
\end{figure*}

The chemical composition of the gas at any point of the simulation, such as the fractional abundance of \CHp, $x(\CHp)$, depends on several local quantities which include the density, $\dens$, the kinetic temperature, $T$, the electronic fraction, $x({\rm e}^-)$, and the molecular fraction\footnote{The molecular fraction is computed as $f(\HH) = 2n(\HH)/\dens$, where $n(\HH)$ is the local density of \HH\ and $\dens$ the local density of protons.}, $f(\HH)$. In addition, the molecular fraction also depends on the shielding factor of \HH\ from the UV dissociative radiation field, $f_{\rm sh, \HH}$ (see Eq. \ref{Eq-H2des}). The regions responsible for the production of \CHp\ in the standard simulation are shown in Fig.~\ref{Fig-form-dist} which displays the 2D probability  distribution functions of the fraction of mass of \CHp\ as a function of $\dens$ and $\dens T$ (top panel), $x(\CHp)$ and $x({\rm e}^-)$ (middle panel), and $f(\HH)$ and $f_{\rm sh, \HH}$ (bottom panel). Fig.~\ref{Fig-form-dist} shows that 90\% of the mass of \CHp\ originates from the warm and unstable gas with $0.6 < \dens < 10$ \cc, $600 < T < 8000$~K, $3 \times 10^{-4} < x({\rm e}^-) < 6 \times 10^{-3}$, $2 \times 10^{-3} < f(\HH) < 0.4$, and $5 \times 10^{-4} < f_{\rm sh, \HH} < 10^{-1}$. The corresponding gas encompasses only $\sim 3.5$\% of the mass of the diffuse ISM and occupies only $\sim 1.5$\% of its total volume. Its physical conditions imply that \CHp\ is primarily destroyed by collision with H\footnote{In particular, the dissociative recombination of \CHp\ is found to have a negligible impact on its chemistry. The abundances of \CHp\ therefore weakly depend on whether the density of electrons is computed at equilibrium or not.}. Yet, its production through the hydrogenation of \Cp\ is so efficient that \CHp\ reaches typical fractional abundances between $3 \times 10^{-9}$ and $4 \times 10^{-6}$. In these regions, \CHp\ can therefore carry as much as $\sim 3$\% of the elemental abundance of Carbon\footnote{The gas phase fractional elemental abundance of Carbon is set to $1.4 \times 10^{-4}$ in the chemical solver.}. Interestingly, we find that the total mass of \CHp\ predicted by the simulation strongly depends on whether the abundance of \HH\ is computed at equilibrium by the chemical solver ($M(\CHp) \sim 3.1 \times 10^{-4}$ M$_\odot$) or out-of-equilibrium during the simulation itself ($M(\CHp) \sim 1.1 \times 10^{-2}$ M$_\odot$). All these behaviors result from the following physical processes.

The steady state obtained in numerical simulations of turbulent and multiphase ISM is statistical in nature (e.g., \citealt{Seifried2011}, \citealt{Saury2014}, \citetalias{Bellomi2020}). The forcing applied at large scale at regular time intervals and the resulting turbulent cascade lead to strong pressure variations and the formation of CNM structures at all scales which are not at thermodynamical equilibrium. Because the volume and the mass of the gas are fixed, the combined actions of the turbulent forcing and the thermal instability induce mass exchanges between the two stable phases, the WNM ($T \sim 8000$ K) and the CNM ($T < 300$ K), thus the existence of a substantial amount of gas in the lukewarm neutral medium (LNM) at intermediate temperatures. Whether this unstable phase triggers the formation of \CHp\ or not depends on the history of the gas. Because the WNM is poor in \HH\ ($f(\HH) \sim 10^{-7}$, see \citetalias{Bellomi2020}), LNM gas transiting from the WNM to the CNM weakly contributes to the production of \CHp. The formation of \CHp\ in the LNM is thus necessarily dominated by the evaporation of CNM clouds, initially rich in $\HH$. This scenario requires that the \HH\ injected in the warm phase survives long enough to allow the production of \CHp\ and that the formation of \CHp\ occurs quickly. Because of the efficient shielding of the surrounding environment, the \HH\ injected in the LNM has a destruction timescale up to $2 \times 10^6$ yr (see bottom panel of Fig.~\ref{Fig-form-dist} and Eq. \ref{Eq-H2des}), a period larger than the dynamical timescale 
($< 10^5$ yr). 
In comparison, the rates given in Table~\ref{tab-reactions} and the conditions of production of \CHp\ shown in Fig.~\ref{Fig-form-dist} imply that \CHp\ reaches its local equilibrium abundance on timescales smaller than $\sim 100$ yr. This short timescale not only insures a maximum production of \CHp\ but also justifies, a posteriori, the use of the chemical solver to postprocess the simulation.


\section{Comparison with observations} \label{Sect-Res}

The maps of the column densities of \HH\ and \CHp\ integrated along the $x$ axis across the standard simulation are shown in Fig.~\ref{Fig-map}. The production of \CHp\ induced by the injection of \HH\ in the warm and diffuse ISM leads to a substantial amount of lines of sight rich in \CHp\ with column densities larger than $10^{12}$ cm$^{-2}$ and up to $\sim 10^{14}$ cm$^{-2}$. The surface filling factor of these lines of sight over a box of 200 pc of diffuse neutral gas is $56\%$. Because the size $L$ of neutral medium is fixed, these distributions of column densities cannot be directly compared with the observations. A quantitative and statistical comparison with the observational data requires to take into account the length of diffuse neutral material intercepted by the observed lines of sight.

\subsection{Distribution of simulated lines of sight}

\begin{figure}[!h]
\begin{center}
\includegraphics[width=9.5cm,trim = 0.7cm 2cm 0.5cm 1.0cm, clip,angle=0]{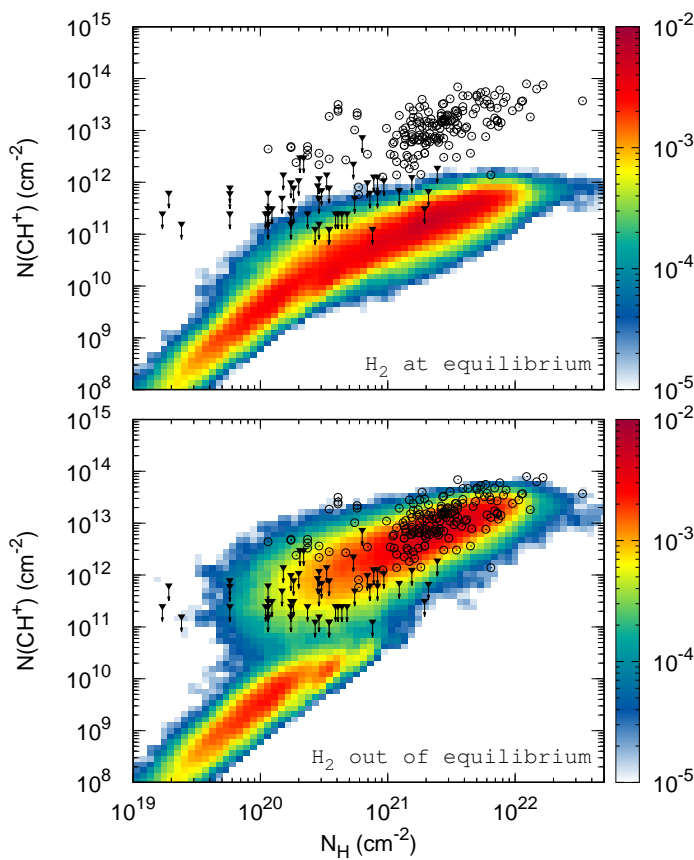}
\caption{Comparison of the observational dataset (black points) with the 2D probability histogram of $N_{\rm H}$ and $N(\CHp)$ predicted by the fiducial simulation (colored histogram) after application of the chemical solver and the lines of sight reconstruction algorithm. The observational dataset include detections (open circles) and non-detections (filled triangles) of \CHp. The color code indicates the fraction of lines of sight (in logarithmic scale) contained in each bin. Two cases are shown: the chemical solver computes the abundance of H$_2$ at equilibrium (top panel) or uses the out-of-equilibrium abundance of H$_2$ calculated during the simulation (bottom panel).}
\label{Fig-fiducial}
\end{center}
\end{figure}

The comparison with the observations is performed statistically, using the methodology proposed in \citetalias{Bellomi2020} (Sect. 3.8 and Fig.~5). We assume that a simulation of size $L$ is a building-block of the neutral diffuse ISM. We construct a random sample of $10^5$ lines of sight with a length distribution given by the distribution of the observed sample (bottom panel of Fig.~\ref{Fig-position-source}). Each line of sight is supposed to intercept diffuse neutral matter but also hot and fully ionized gas. The fraction of the line of sight occupied by this hot ionized medium is set to a constant value of $0.5$ (see Sect. 5.3 of \citetalias{Bellomi2020}). The total proton column density, $N_{\rm H}$, and the column density of \CHp, $N(\CHp)$, along each line of sight are built by integrating the abundances along the $x$ axis\footnote{Because the magnetic field energy density is at equipartition with the kinetic and thermal energy densities, integrations performed along the $y$ and $z$ axis lead to identical statistical results.} over a fraction of the simulation if the length occupied by the neutral material is smaller than $L$, or by stacking together several random lines of sight of size $L$ otherwise.

The resulting 2D probability histograms of $N_{\rm H}$ and $N(\CHp)$ are shown and compared with the observations in Fig.~\ref{Fig-fiducial}. The color code indicates the fraction of simulated lines of sight with a given chemical composition ($N_{\rm H}$,$N(\CHp)$). The results shown here correspond to the fiducial simulation (Table~\ref{tab-params}) which is found to display an outstanding statistical agreement with the observations of the HI-to-\HH\ transitions (see Fig. \ref{Fig-main-prop}). If the abundance of \HH\ is computed at equilibrium, the simulation underestimates the column densities of \CHp\ by about two orders of magnitude (top panel of Fig.~\ref{Fig-fiducial}). Yet, if \HH\ is computed out-of-equilibrium, the very same setup appears to naturally reproduce the distribution of observations of $N_{\rm H}$ and $N(\CHp)$ (bottom panel of Fig.~\ref{Fig-fiducial}). 
For $N_{\rm H} > 10^{21}$ cm$^{-2}$, the simulation predicts that most of the lines of sight should have large column densities of \CHp, with  ${\rm log}[ N(\CHp)/N_{\rm H}] \sim -8.57 \pm 0.28$ in remarkable agreement with the mean and dispersion values of the observational sample in this domain ($-8.37 \pm 0.32$). For $N_{\rm H} < 10^{21}$ cm$^{-2}$, the simulated and observed samples must be compared by taking into account the number of non-detections. The observations show a limit of detection of \CHp\ of $\sim 10^{12}$ cm$^{-2}$. For $N_{\rm H} <10^{20}$ cm$^{-2}$, 98\% of the simulated lines of sight have a column density of \CHp\ below $10^{12}$ cm$^{-2}$ in agreement with the number of non-detections in this domain (100\%). For $10^{20}< N_{\rm H} <10^{21}$ cm$^{-2}$, the predicted and observed fractions of non-detections similarly drop to 60\% and 65\% respectively. In this domain, the simulated lines of sight with $N(\CHp) > 10^{12}$ cm$^{-2}$ verify ${\rm log}[ N(\CHp)/N_{\rm H}] \sim -8.27 \pm 0.32$, in agreement, although slightly lower than the observed values ($-7.85 \pm 0.54$).

It is important to note that the mass flow rate between the CNM and the WNM induced by large scale turbulence not only produces an amount of \CHp\ comparable to the observations but also naturally generate all the statistical properties seen in the observations. As shown in Fig.~\ref{Fig-fiducial}, this process simultaneously explains the probabilities of occurrence of most of the lines of sight, the probabilities of non detections of \CHp, the mean fractional abundances of \CHp\ and their dispersions, all as functions of the total proton column density $N_{\rm H}$. Such a precise description of the observational data has never been achieved, nor even been searched for, by any previous theoretical model.

In the following section, we deepen the analysis by comparing the results of the simulation to another paramount, yet poorly studied, observational signature: the kinematic information carried in the line profiles of \CHp.

\subsection{Distribution of line profiles}

\begin{figure*}[!h]
\begin{center}
\includegraphics[width=18.0cm,trim = 2.0cm 6.5cm 0.3cm 0.5cm, clip,angle=0]{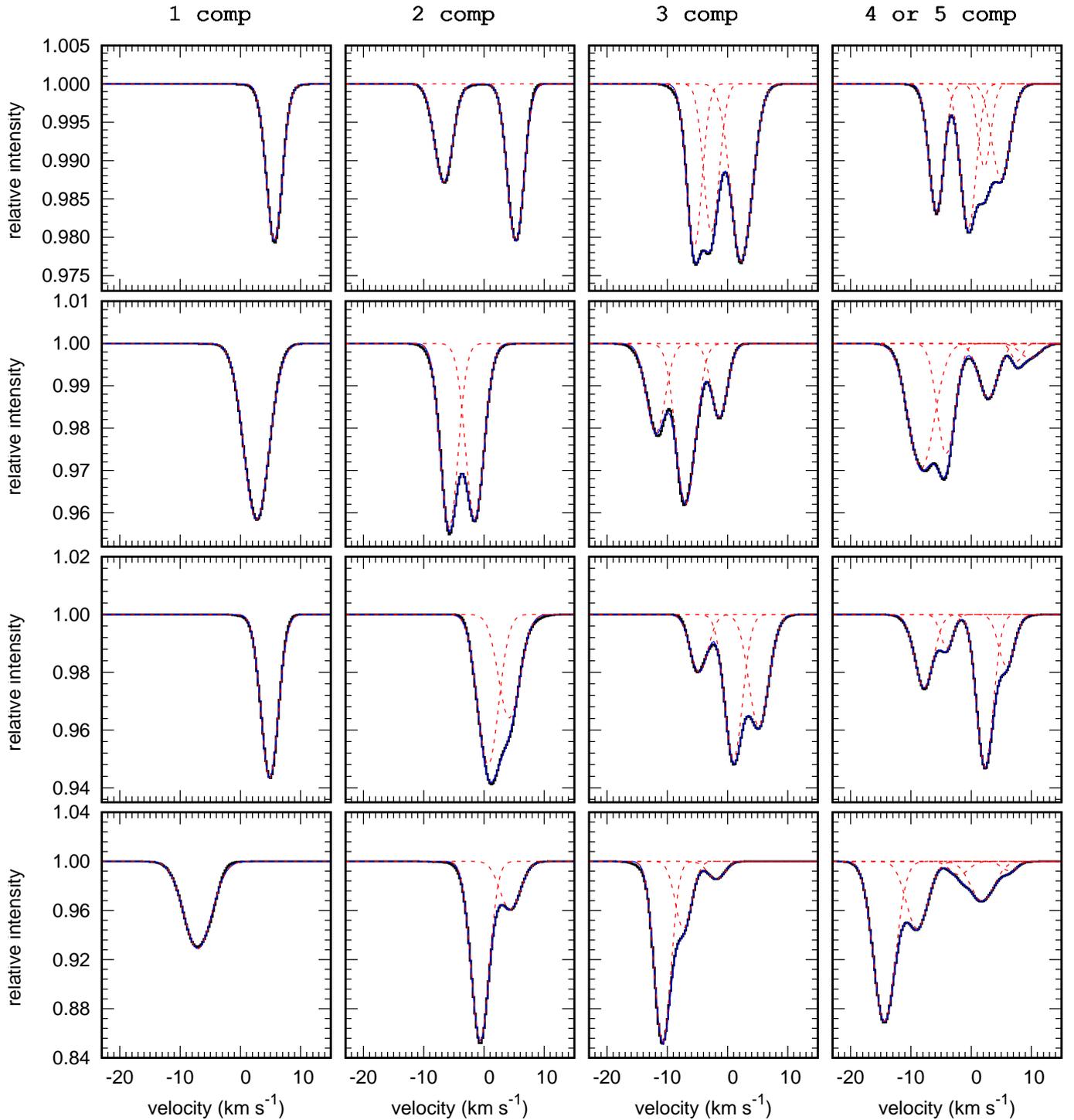}
\caption{Examples of absorption profiles of the 4232 \AA\ line of \CHp\ computed along the $x$ axis across the standard simulation (see Table~\ref{tab-params}). The black histograms are the computed spectra. The colored curves are the results of a multiGaussian fit applied to the line opacity and display the individual Gaussian components (dashed red curves) and their sum (solid blue curves). The spectra are classified in increasing order of their maximal opacity (from top to bottom) and in increasing order of the number of Gaussian components used in the fit (from left to right).}
\label{Fig-simprof}
\end{center}
\end{figure*}

\begin{figure}[!h]
\begin{center}
\includegraphics[width=9.5cm,trim = 1.5cm 2cm 0.5cm 1.0cm, clip,angle=0]{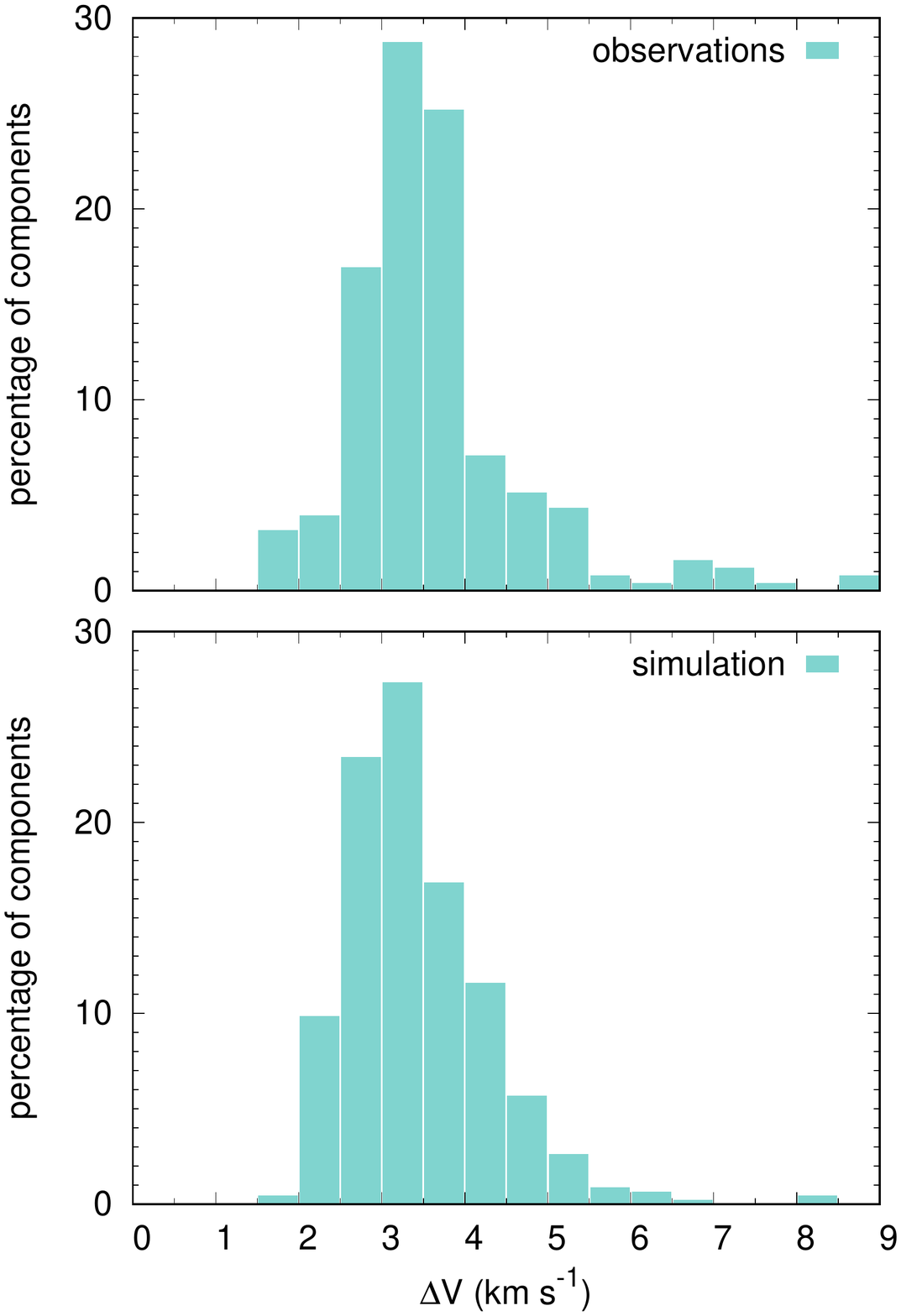}
\caption{Probability distributions of the full width at half maximum $\Delta V$ of the 4232 \AA\ line of \CHp\ derived from the observations (top panel) and from synthetic spectra (see Fig.~\ref{Fig-simprof}) computed across the standard simulation (bottom panel). The observational sample includes 254 velocity components extracted along 120 lines of sight by \citet{Federman1982} (Table~2), \citet{Lambert1986} (Table~1), \citet{Crane1995} (Table~2), \citet{Pan2004} (Table~4), \citet{Ritchey2006} (Table~4), and \citet{Sheffer2008} (Table~3). This dataset includes all the lines of sight where \CHp\ has been observed (see Table~\ref{Tab-Obs} of Appendix \ref{AppendObs}), except those observed or compiled by \citet{Gredel2002} and \citet{Rachford2002} who do not provide measurements of the \CHp\ linewidth.}
\label{Fig-distdV}
\end{center}
\end{figure}

The observations performed in the optical and submillimetre domains have long revealed the kinematic signatures of \CHp\ (e.g., \citealt{Crane1995}). MultiGaussian fits of the observations systematically show that the \CHp\ absorption lines have typical linewidths of a few \kms\ and are broader than the absorption lines of most of the molecules and molecular ions observed in the diffuse ISM (e.g., \citealt{Crane1995,Godard2012}). In fact, principal component analysis of absorption spectra show that the kinematic profile of \CHp\ is similar to that of HI \citep{Neufeld2015}. Such broad absorption profiles are a firm signature of the regions that produce \CHp. Because they follow the 3D dynamical, thermal, and chemical states of the gas, the simulations used in this work open the possibility to perform a comparative study between observational and theoretical line profiles. This analysis is done by applying the following methodology.

Synthetic spectra of the $X^1\Sigma^+ - A^1\Pi$ 4232 \AA\ line of \CHp\ are computed across the standard simulation along 200 random lines of sight with $N(\CHp) > 10^{12}$ cm$^{-2}$, a value chosen from the observational detection limit of \CHp\ (see Fig.~\ref{Fig-fiducial}). Because the density and the kinetic temperature are too low to significantly excite the $A^1\Pi$ state of \CHp, the emissivity of this line is negligible along the line of sight. The velocity dependent specific intensity $I(\upsilon)$ relative to that of the continuum background $I_c$ is thus simply calculated as
\begin{equation}
\frac{I(\upsilon)}{I_{c}} = {\rm exp} \left( -\tau(\upsilon) \right).
\end{equation}
The line opacity $\tau(\upsilon)$ is integrated along the $x$ axis
\begin{equation}
\tau(\upsilon) = \left( \frac{c}{\nu_0} \right)^2 g_u \frac{A_{ul}}{8\pi}  \int_0^L n(\CHp) \, \phi(\upsilon) \, dx,
\end{equation}
where $\nu_0$, $A_{ul}$, and $g_u$ are the line rest frequency, its spontaneous emission rate, and the degeneracy of the upper level, $c$ is the speed of light, and $n(\CHp)$ is the local density of \CHp. The local line profile $\phi(\upsilon)$ is assumed Gaussian and computed as
\begin{equation}
\phi(\upsilon) = \frac{1}{\sqrt{2 \pi} \sigma}  {\rm exp} \left( - \frac{1}{2} \left[ \frac{\upsilon - \upsilon_0}{\sigma} \right]^2 \right)
\end{equation}
where $\upsilon_0$ is the local velocity of the gas projected along the $x$ axis and $\sigma$ is the local 1D velocity dispersion
\begin{equation}
\sigma = \left( \frac{kT}{m} + \sigma_{\rm tur}^2 \right)^{1/2}
\end{equation}
set by the local kinetic temperature, $T$, the mass of \CHp, $m$, and the micro-turbulent velocity dispersion $\sigma_{\rm tur}$ at the resolution scale of the simulation. Because $\sigma_{\rm tur}$ is, by construction, inaccessible from the simulation, its value is set from the velocity dispersion-size relation of molecular clouds (e.g., \citealt{Larson1981,Hennebelle2012}) as
\begin{equation}
\sigma_{\rm tur} = 1 \kms\ \left( \frac{L / {\rm pc}}{R^{1/3}} \right)^{1/2},
\end{equation}
where $L$ is the size of the simulated box expressed in pc and $R$ is the numerical resolution. Such a choice implies that the micro-turbulent velocity dispersion of the standard simulation, $\sigma_{\rm tur} = 0.625$ \kms, is always smaller than the thermal velocity dispersion of the gas producing \CHp\ ($T>600$ K, see Fig.~\ref{Fig-form-dist}) and has therefore a weak impact on the resulting line profiles.

Once computed, the 200 synthetic spectra are analyzed with the same method than that applied to real observations. The opacity profile of each spectrum is decomposed into velocity components by performing a multiGaussian fit using the curve fit optimization algorithm provided by the SciPy python library. The number of components is chosen as the minimal number of Gaussian features required to reproduce the spectrum with residual opacities about 10 times below the peak opacity of the spectrum. 
Examples of the synthetic spectra obtained across the standard simulations and their multiGaussian decompositions are displayed in Fig.~\ref{Fig-simprof}. The detection limit $N(\CHp)> 10^{12}$ cm$^{-2}$ implies typical peak opacities of the 4232 \AA\ line of \CHp\ larger than $\sim 0.01$. The distribution of phases along the lines of sight, the large scale velocity dispersion of the diffuse ISM set by the turbulent forcing, and the velocity dispersion of the gas producing \CHp, frequently lead to blended features that require more than one Gaussian component. 24\%, 38\%, 29\%, 8\%, and 2\% of the spectra  across 200 pc of diffuse neutral medium are fitted with one, two, three, four, and five Gaussian components respectively. In total, the 200 synthetic spectra are decomposed into 457 velocity components.

The resulting distribution of the full width at half maximum $\Delta V$ of the velocity components of \CHp\ is shown in Fig.~\ref{Fig-distdV} (bottom panel) and compared to the distribution extracted from optical observations of \CHp\ in the local ISM (top panel). Surprisingly, the thermodynamical properties of the turbulent multiphase ISM obtained in the standard simulation are found to naturally explain the observed distribution of the \CHp\ absorption profiles. Most of the velocity components of \CHp\ are predicted to have linewidths between 1.5 and 5 \kms, with a tail up to $\sim 10$ \kms, as seen in the observational sample. The distribution obtained with the numerical simulation leads to an expected linewidth $\Delta V \sim 3.44 \pm 0.87$ \kms\ in agreement with the mean and dispersion values of the observational dataset ($3.59 \pm 1.18$ \kms).

The results displayed in Fig.~\ref{Fig-distdV} are promising and open new prospects for statistical comparisons between simulations and observations. It would be tempting, for instance, to perform a more advanced analysis of these distributions including the number of velocity components per line of sight and the probability of occurrence of each linewidth. Such a detailed comparison requires, however, to slightly adapt our methodology. Indeed, (1) the fact that the simulated spectra are built without noise necessarily leads to more and narrower velocity components than those extracted from real observational data. Moreover, (2) a detailed comparison with the observations would require to compute synthetic spectra along lines of sight that follow the length distribution of the observed sample (bottom panel of Fig.~\ref{Fig-position-source}) as was done above for the comparisons of the column densities. Finally, (3) the micro-turbulent velocity dispersion (which has no impact in the current predictions) should be described more carefully. On the one side, and because of numerical diffusion, the micro-turbulent velocity is probably underestimated and should be derived at slightly larger scales than the numerical resolution scale. On the other side, this micro-turbulent velocity dispersion is deduced from the velocity dispersion-size relation of molecular clouds which has no reason to apply to the evaporating medium responsible for the production of \CHp. All these considerations are beyond the scope of the present paper.

\section{Exploration of the parameters} \label{Sect-params}

In this section, we discuss the dependence of the production of \CHp\ on several key parameters including the resolution of the simulation $R$, the mean density $\overline{\dens}$, the UV illumination factor $G_0$, the strength of the turbulent forcing $F$, and its compressive ratio $\zeta$. The exploration is performed over a set of 26 simulations. Each run is postprocessed with the chemical solver using the out-of-equilibrium abundance of \HH\ computed during the simulation. A simulated sample of lines of sight is then drawn following the distribution of length of the observed sample. The influence of the parameters are estimated and discussed based on their impact on the probability histograms of $N_{\rm H}$ and $N(\CHp)$ compared to that of the fiducial setup (Fig.~\ref{Fig-fiducial}).

\subsection{Influence of the resolution}

\begin{figure}[!h]
\begin{center}
\includegraphics[width=9.0cm,trim = 1.5cm 2.0cm 1.5cm 2.0cm, clip,angle=0]{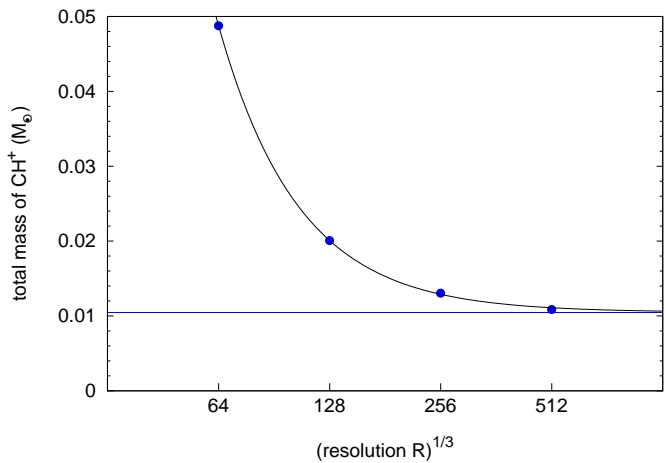}
\caption{Total mass of \CHp, $M_{\rm tot}(\CHp)$, contained in simulations at different resolutions $R$ (blue points). The data are fitted with the function $a / R^{b/3} + c$, where $a$, $b$, and $c$ are adjustable coefficients. The best fit (black curve) corresponds to the function $1.57 \times 10^{2} / R^{2/3} + 1.05 \times 10^{-2}$ and tends towards the constant value $1.05 \times 10^{-2}$ (blue line) at high resolution.}
\label{Fig-massCHp-reso}
\end{center}
\end{figure}

\begin{figure}[!h]
\begin{center}
\includegraphics[width=9.8cm,trim = 0.0cm 2cm 2.0cm 1.5cm, clip,angle=0]{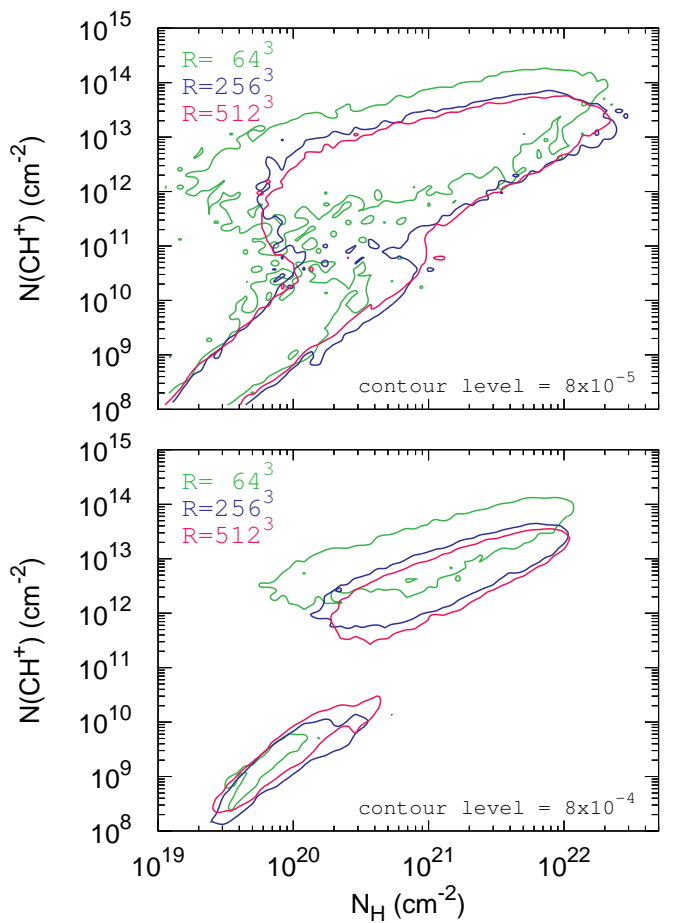}
\caption{Isocontours of the 2D probability histograms of $N_{\rm H}$ and $N(\CHp)$ predicted by simulations with a resolution $R=64^3$ (green curve), $256^3$ (blue curve), and $512^3$ (pink curve), after application of the chemical solver and the lines of sight reconstruction algorithm. Such representation shows, in a simplified manner, the effect of the resolution on the result displayed in Fig. \ref{Fig-fiducial}. The isocontours are chosen at $8 \times 10^{-5}$ (top panel) and $8 \times 10^{-4}$ (bottom panel) to  highlight the changes in the distribution of the most probable lines of sight. These values correspond to the turquoise blue and orange levels of the 2D probability histograms of $N_{\rm H}$ and $N(\CHp)$ (see Fig.~\ref{Fig-fiducial}).}
\label{Fig-grid-reso}
\end{center}
\end{figure}

In \citetalias{Bellomi2020}, we found that the HI-to-\HH\ transition is independent of the resolution over about one order of magnitude ($64^3 \leqslant R \leqslant 512^3$). This is understood by the fact that \HH\ is almost entirely produced in the CNM. The predicted distribution of the column densities of \HH\ therefore only depends on whether the simulation accurately resolves the CNM structures that carry most of the mass and volume of \HH. While simulations at high resolution allow the formation of dense structures at small scales, those have a negligible impact on the mass of \HH\ and the volume it occupies. 
Such a result is expected to hold for any other chemical tracer as long as it is formed in the same regions than \HH. This is not the case of \CHp.

The impact of the resolution on the production of \CHp\ is shown in Fig.~\ref{Fig-massCHp-reso} which displays the total mass of \CHp contained in simulations at different resolutions, and in Fig.~\ref{Fig-grid-reso} which compares two isocontours of the 2D probability histograms of $N_{\rm H}$ and $N(\CHp)$ for $R=64^3$, $256^3$, and $512^3$. Oppositely to the HI-to-\HH\ transition, Figs.~\ref{Fig-massCHp-reso} and \ref{Fig-grid-reso} shows that the distribution of the column densities of \CHp\ strongly depends on the resolution for $R < 256^3$ and is almost independent of the resolution for $R \geqslant 256^3$. Quantitatively, the mass of \CHp\ produced in the simulation at $R=64^3$ is $\sim 5$ times larger than that produced for $R=512^3$. In comparison, the simulation at $R=256^3$ produces only $\sim 20\%$ more \CHp\ than the simulation at $R=512^3$. This behavior, which significantly differs from that of the HI-to-\HH\ transition, is due to the origin of \CHp. While \HH\ is built in the CNM, \CHp\ is  entirely produced in the unstable diffuse ISM (LNM). As described in Sect. \ref{Sect-origin}, the total mass of \CHp\ in these regions not only depends on their physical conditions (density, temperature, electronic fraction, and molecular fraction) which set the local abundance of \CHp, but also on the rate at which \HH\ is injected from the CNM to the LNM. Interestingly, the analysis of the simulations at different resolutions show that the physical conditions of these regions are independent of the resolution. The observed dependence therefore solely results from changes of the rate of injection of \HH\ in the LNM. The results displayed in Figs.~\ref{Fig-massCHp-reso} and \ref{Fig-grid-reso} suggest that this injection rate is dominated by numerical diffusion for $R < 256^3$ and by a real physical process for $R \geqslant 256^3$. The nature of this process is discussed in Sect. \ref{Sect-turb}.

The weak dependence of the 2D probability histograms of $N_{\rm H}$ and $N(\CHp)$ for a resolution $R \geqslant 256^3$ indicates that the numerical convergence is reached at large resolution. Moreover, it justifies the fact to use simulations with a moderate numerical resolution to study the production of \CHp\ in the diffuse interstellar medium. The exploration of the parameter domain in the following sections are therefore performed for $R = 256^3$.


\subsection{Impact of $G_0$ and $\overline{\dens}$}

\begin{figure*}[!h]
\begin{center}
\includegraphics[width=16.5cm,trim = 1.3cm 2cm 0.5cm 3.0cm, clip,angle=0]{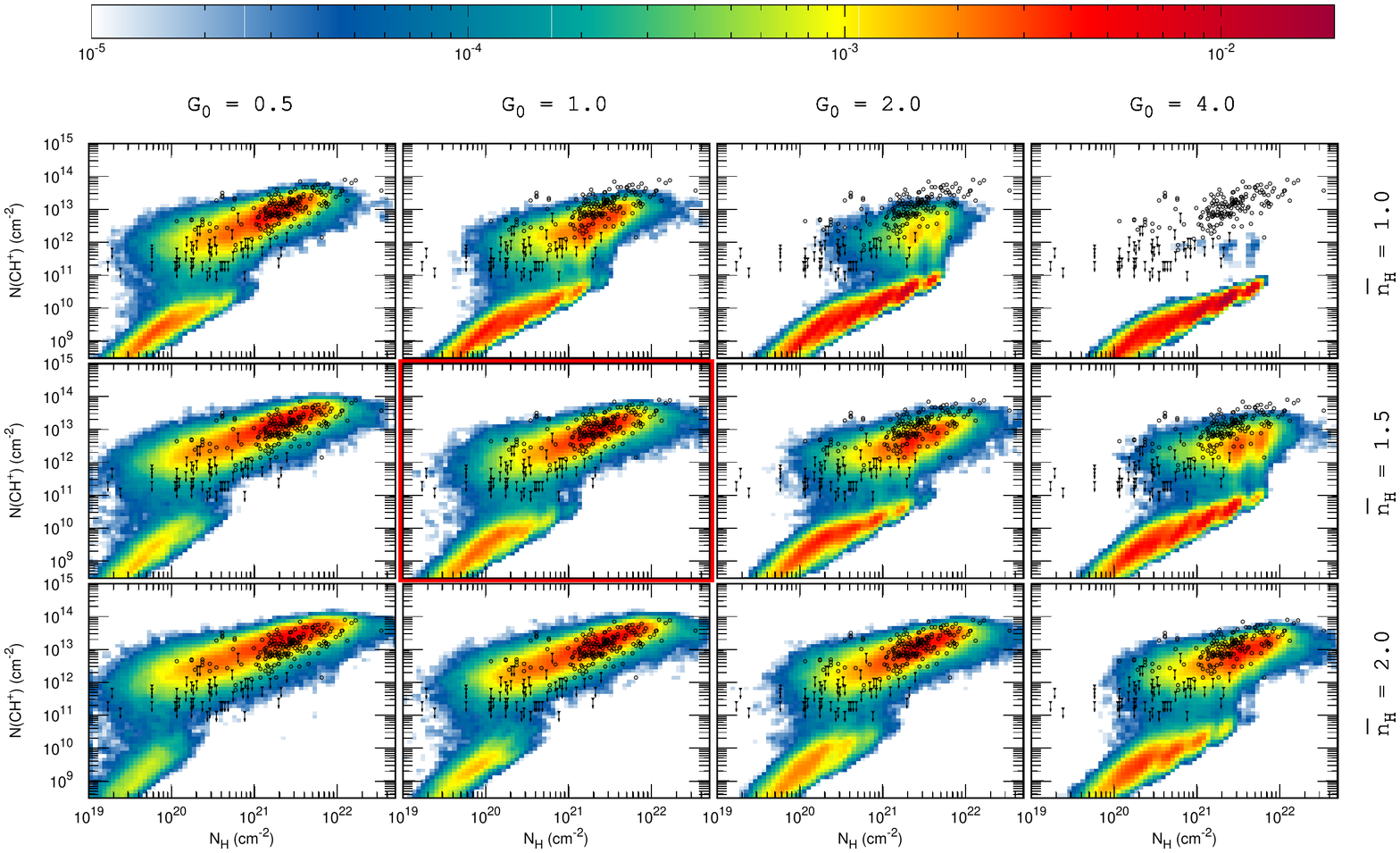}
\includegraphics[width=16.5cm,trim = 1.3cm 2cm 0.5cm 5.0cm, clip,angle=0]{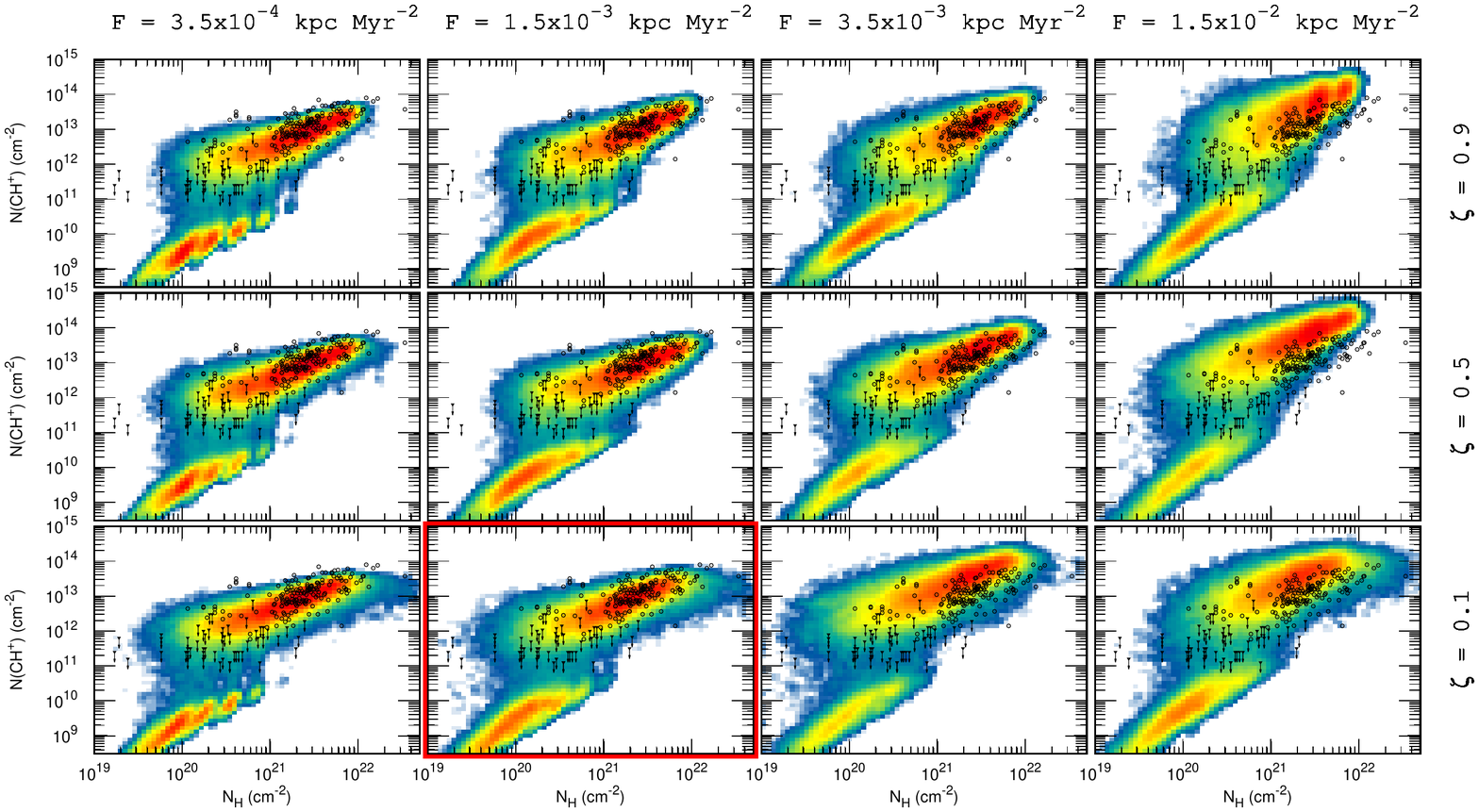}
\caption{Comparison of the observational dataset (black points) with the 2D probability histograms of $N_{\rm H}$ and $N(\CHp)$ predicted by numerical simulations (colored histograms) after application of the the chemical solver and the lines of sight reconstruction algorithm. The observational dataset include detections (open circles) and non-detections (filled triangles) of \CHp. Each panel in the top frame shows the prediction of a different simulation with $G_0$ varying between 0.5 and 4 (from left to right) and $\overline{\dens}$ varying between 1 and 2 \cc\ (from top to bottom). Each panel in the bottom frame shows the prediction of a different simulation with $F$ varying between $3.5 \times 10^{-4}$ and $1.5 \times 10^{-2}$ kpc Myr$^{-2}$ (from left to right) and $\zeta$ varying between 0.9 and 0.1 (from top to bottom). If not indicated otherwise, all the other parameters are set to their fiducial values (see Table~\ref{tab-params}), except for the resolution which is set here to $R=256^3$. The color code indicates the fraction of lines of sight (in logarithmic scale) contained in each bin. The simulation corresponding to the standard set of parameters (except for the resolution) is highlighted with a red square.}
\label{Fig-grid}
\end{center}
\end{figure*}

The impacts of the mean density and the strength of the UV radiation field are displayed in Fig.~\ref{Fig-grid} (top frame), which shows comparisons of the 2D probability histograms of $N_{\rm H}$ and $N(\CHp)$, predicted for various values of $\overline{\dens}$ and $G_0$, to the observational sample. The results obtained here are a direct consequence of those displayed in Fig.~9 of \citetalias{Bellomi2020} which shows the dependence of the HI-to-\HH\ transition on these two parameters over a larger range of mean density $\overline{\dens}$. Indeed, since the amount of \CHp\ is proportional to the rate of injection of \HH\ in the LNM, \CHp\ is tightly linked to the reservoir of \HH\ itself. An (under)overestimation of the column densities of \HH\ in the local diffuse ISM directly translates into an (under)overestimation of the column densities of \CHp. Interestingly, we find that the statistical distribution of the observed column densities of \CHp\ can only be reproduced if the statistical distribution of the observed column densities of \HH\ are reproduced as well (see Fig. \ref{Fig-main-prop}). These separated contraints emphasize the coherence of the chemical composition predicted with the standard setup. 

As explained in \citetalias{Bellomi2020}, the strong dependences of the results on $\overline{\dens}$ and $G_0$ only weakly originate from changes of the local physical conditions. They are in fact a consequence of the statistical distributions of phases along random lines of sight. The thermal instability applied to diffuse gas with a constant volume and a constant mass implies that decreasing $\overline{\dens}$ or increasing $G_0$ considerably reduce the mass and the volume occupied by of the cold and dense CNM. This not only reduces the total mass of \HH\ but also the probability to obtain lines of sight with large molecular fraction. Such lines of sight therefore occur at larger total column density $N_{\rm H}$ and their molecular fractions exhibit a larger dispersion around the mean value. As expected from the interpretative framework given in Sect. \ref{Sect-origin} and as shown in the top frame of Fig.~\ref{Fig-grid}, the column densities and the total mass of \CHp\ reveal the exact same dependences as the molecular fraction and the total mass of \HH.

Figs.~\ref{Fig-main-prop} and \ref{Fig-fiducial} indicate that the fiducial setup slightly overestimate the mean pressure of the ISM, and slightly underestimate the total mass of \HH\ and \CHp\ (by a factor $\sim 1.5$). As shown in Fig. \ref{Fig-grid}, these shallow discrepancies could be solved by slightly adjusting the mean density (above 1.5 \cc) or the impinging radiation field (below 1). We consider such fine tuning of the model parameters superfluous given the uncertainties on the key underlying microphysical processes discussed in Sect. \ref{Sect-Dis} and on the turbulent forcing.

\subsection{Impact of $F$ and $\zeta$} \label{Sect-turb}

The strength and the nature of the turbulent forcing are known to have a significant influence on the physics of the multiphase ISM (e.g., \citealt{Seifried2011}). The variations of pressure and the shear motions induced at all scales by the turbulent forcing not only perturb the gas out of its thermal equilibrium states, but also reduce the time spent in the WNM, LNM, and CNM \citep{Seifried2011}. On the one side, the turbulent forcing thus modifies the mass and volume distributions of the phases: increasing the strength of the turbulent forcing leads to a diminution of the mass of the CNM to the benefit of the LNM and the WNM ; this effect is, however, much larger for a solenoidal forcing than a compressive forcing because solenoidal motions efficiently prevent the gas to condensate back to the CNM (see Fig.~4 and Sect. 3.7 of \citetalias{Bellomi2020}). On the other side, the turbulent forcing modifies the rate at which mass is exchanged between the different phases: increasing the strength of the turbulent forcing leads to an increase of the mass transfer rate between the CNM, the WNM, and the LNM.

The impact on the HI-to-\HH\ transition is presented and discussed in Sect. 4.6 of \citetalias{Bellomi2020}. In a nutshell, the HI-to-\HH\ transition is found to weakly depend on the strength of the turbulent forcing if the forcing is dominated by compressive motions, and to be sensitive to the strength of the forcing if the forcing is dominated by solenoidal motions. Overall, the turbulent forcing has a moderate impact because the HI-to-\HH\ transition is only sensitive to the distributions of mass and volume occupied by the different phases. This is not the case of \CHp\ which also depends on the mass loss rate of the CNM.

The influence of the strength and the nature of the turbulent forcing on the production of \CHp\ are displayed in Fig.~\ref{Fig-grid} (bottom frame) which shows comparisons of the 2D probability histograms of $N_{\rm H}$ and $N(\CHp)$, predicted for various values of $F$ and $\zeta$, to the observational sample. The dependence of \CHp\ on the turbulent forcing is significantly different from that of the HI-to-\HH\ transition. 
For a kinetic energy density (which is roughly proportional to $F$) above the thermal energy density ($F \geqslant 1.5 \times 10^{-3}$ kpc Myr$^{-2}$, see Table \ref{tab-main-prop}), the distribution of the column densities of \CHp\ is found to strongly depend on the strength of the turbulent forcing regardless of its compressive or solenoidal nature.  Increasing the strength of the turbulent forcing from $F=1.5 \times 10^{-3}$ kpc Myr$^{-2}$ to $F=1.5 \times 10^{-2}$ kpc Myr$^{-2}$ leads to an increase of the total mass of \CHp\ by about an order of magnitude. The amplitude of this effect depends on the nature of the forcing  and is slightly smaller if the forcing is dominated by solenoidal motions. All these behaviors confirm the interpretation given in Sect. \ref{Sect-origin}. The amount of \CHp\ is proportional to the rate of injection of \HH\ in the LNM. This rate increases with the strength of the turbulent forcing (if the kinetic energy density is larger than the thermal energy density) and slightly decreases with the fraction of power injected in the solenoidal modes which efficiently reduce the mass of the CNM. Interestingly, the tightest agreement with the observations of \CHp\ and the HI-to-\HH\ transition is obtained for a forcing which is not purely solenoidal ($\zeta \leqslant 0.5$) with $F \sim 1.5 \times 10^{-3}$ kpc Myr$^{-2}$. We recall that this forcing induces a 1D turbulent velocity dispersion of the WNM of $\sim 5.4$ \kms\ (see Sect. \ref{Sect-standard}), in agreement with the velocity dispersion derived from HI emission spectra at high Galactic latitude \citep{Kalberla2005,Haud2007}.

Interestingly, Fig. \ref{Fig-grid} indicates that the turbulent forcing has a weak impact on the production of \CHp\ if the kinetic energy density is below or close to the thermal energy density ($F \leqslant 1.5 \times 10^{-3}$ kpc Myr$^{-2}$). This result should, however, be considered with caution. Indeed, study of the numerical convergence was performed for $F = 1.5 \times 10^{-3}$ kpc Myr$^{-2}$ (see Figs.~\ref{Fig-massCHp-reso} and \ref{Fig-grid-reso}). While convergence is expected to be reached for larger values of $F$, it is not necessarily the case for lower values where the evaporation process of CNM clouds could be mediated by artificial thermal conduction. Confirmation of this result would require to perform a convergence study for smaller values of the strength of the turbulent forcing or even without forcing where the turbulence is sustained by the thermal instability itself \citep{Iwasaki2014}.

\section{Discussion} \label{Sect-Dis}

The simulations presented in this work show that it is possible to build a coherent physical model of the multiphase diffuse ISM that simultaneously explains a great variety of observations, including chemical correlations, statistical distributions, and line profiles, within the constraints set by the structure of the ISM observed at Galactic scales (e.g., ISM midplane mean density, UV interstellar radiation field, and illumination scales). We call this model the standard simulation. It is important to note, however, that such a model is not unique and depends on the prescriptions used for several microphysical processes. Any uncertainty on these processes will likely result in an update of the so-called standard simulation.

The mean pressure of the diffuse ISM depends on the efficiency of the photoelectric heating (e.g. \citealt{Wolfire2003}). This efficiency, in turn, depends on the abundance of very small grains, on their recombination rate with free electrons, on the photoelectric yield, and on their very nature including their composition and structure. \citet{Weingartner2001a} estimate that the uncertainties on the yield and the recombination rate alone implies an uncertainty on the photoelectric heating rate of about a factor of 3. In this paper, we model very small grains as carbonaceous PAHs, with a fractional abundance of $10^{-6}$ and assume a sticking coefficient of free electrons of 1. Such choices lead to a photoelectric efficiency about twice as large as the reference model of \citet{Weingartner2001a}. Other prescriptions for the photoelectric effect could be adopted but would require to modify the standard setup in order to account for the observed mean pressure of the CNM.

Although not displayed in this paper, the effect of the initial homogeneous magnetic field has been explored. As long as the magnetic energy density is below equipartition, this parameter is found to have a small influence on the distribution of pressure and the statistical distribution of the HI-to-\HH\ transition. A reduced strength of the magnetic field leads, however, to an increase of the total mass of \CHp\ because it facilitates the exchanges of mass between the CNM and the WNM. We find that reducing the magnetic field intensity by a factor of 5 leads to an amount of \CHp\ twice as large as in the standard setup.

The column densities predicted by the simulations obviously depend on the chemical rates adopted for the formation and the destruction of \CHp. We note that the rates used in this work (see Table~\ref{tab-reactions}) are significantly different from those used in other theoretical studies \citep{Myers2015,Moseley2021}. The fact that both the formation rate of \CHp\ by hydrogenation of \Cp, and the destruction rate of \CHp\ by reactive collision with H are larger than those adopted by \citet{Myers2015} by a factor of 5 has evidently no impact on our results but stresses the need for reliable chemical rates. A more robust treatment would be to take into account the impact of the local distribution of \HH\ in its excited levels in the computation of the formation rate of \CHp\ \citep{Agundez2010,Zanchet2013}, which would lead to a lower formation rate than that prescribed by \citet{Hierl1997}. In contrast, we note that the destruction rate of \CHp\ by collision with H prescribed by \citet{Plasil2011} and adopted in this work appears to overestimate the rate measured at high temperature ($T> 800$ K) by \citet{Federer1984,Federer1985}. We find that implementing the state-to-state chemical rates for the formation of \CHp\ and modifying the destruction of \CHp\ by collision with H to match the measurements performed at high temperature reduce the amount of \CHp\ by a factor two only, i.e., well within the uncertainty expected, given the loose contraints on the strength of the turbulent forcing and the strength of the magnetic field in the local diffuse ISM.

\section{Conclusions} \label{Sect-Con}

The large abundances and spatial coverage of \CHp\ has long been recognized as a deep mystery of the chemical state and evolution of the diffuse interstellar matter. In this paper we explore the impact of the multiphase and turbulent nature of the ISM on the production of \CHp\ and on its kinematic signatures through detailed comparisons of the results of numerical simulations with observations of \CHp\ in the Solar Neighborhood. Because of the long chemical timescale of \HH\ and the short chemical timescale of \CHp, the chemical composition of the ISM is modeled by solving the out-of-equilibrium abundance of \HH\ during the simulation itself and by computing the rest of the chemistry at equilibrium in post-processing.

A first and important outcome of this work is the validation of the method proposed in \citetalias{Bellomi2020} and reapplied here to confront numerical simulations and observations. The validity of the results of a model is estimated from its capacity to account for the full statistical information contained in an observational sample which not only include correlations between the column densities of different species, but also the rate of non-detections of these species, and the probability of occurrence of any groups of lines of sight. On the one side, this detailed and quantitative comparison requires to take into account the distribution of lengths of the intercepted diffuse matter for the construction of a simulated sample of lines of sight. 
On the other side, the method requires that the observed dataset is unbiased. In particular, it relies on the fact that the observational sample corresponds to random lines of sight which are not selected depending on the opacity of the foreground medium. Moreover, it relies on the fact that the information is complete and that upper limits are systematically given if the species is not detected. As far as we know, both conditions are fulfilled in the observational sample of \CHp\ studied here.

\CHp\ is usually considered as a tracer of the intermittent dissipation of turbulence in regions with large ion-neutral velocity drift. The present paper shows, for the first time, that the production of \CHp\ might be a natural consequence of the exchanges of mass between the CNM and the WNM induced by the combination of large scale turbulence and thermal instability. Indeed, the results of a single simulation of the turbulent multiphase ISM 
are able to reproduce most of the statistical properties of the observations of \CHp\ with a precision never achieved before by any theoretical model. In particular, the model explains the probabilities of occurrence of most of the lines of sight, the probabilities of non detections of \CHp, the range of observed column densities, the mean fractional abundances of \CHp\ and their dispersions, all as functions of the total proton column density along the lines of sight (Fig.~\ref{Fig-fiducial}). In addition, the combined dynamical and thermochemical properties of the regions responsible for the production of \CHp\ lead to a distribution of line profiles almost identical to that derived from the observational data (Fig.~\ref{Fig-distdV}). All these results are obtained without tuning any parameter, but by simply using the setup found to reproduce the statistical properties of the HI-to-\HH\ transition and the distribution of thermal pressure observed in the local diffuse ISM. These separated chemical and physical constraints 
highlight the remarkable coherence of the chemical composition obtained with simulations of the turbulent multiphase ISM.

The quasi-totality of the mass of \CHp\ originates from the warm an unstable gas produced by the evaporation of CNM clouds which injects out-of-equilibrium \HH\ in warmer and more diffuse environments. This process depends on the mass and volume distributions of the CNM, primarily set by the mean density of the diffuse ISM and the intensity of the UV radiation field, and the evaporation timescale of the CNM set by the strength of the turbulent forcing. The physical conditions (i.e., the density, the kinetic temperature, the molecular fraction an the electronic fraction) of the regions responsible for the production of \CHp\ weakly depend on the model parameters. The amount of \CHp\ in the diffuse ISM is therefore a sole tracer of the injection rate of \HH\ in the unstable gas.

This injection rate can be roughly estimated using the following argument. Let's consider a volume of CNM in expansion. Since its dynamical timescale is smaller than the destruction timescale of \HH, the mass of \HH\ contained in the volume is initially conserved. The expansion reduces, however, the self-shielding of \HH\ and its destruction timescale. The mass of \HH\ is thus conserved as long as the dynamical timescale is shorter than the chemical timescale, and then exponentially decreases as a function of time. We estimate that this turning point happens for a self-shielding factor $f_{\rm sh, \HH} \sim 10^{-2}$. It follows that the total mass of warm \HH\ ($T>600$ K) contained in the standard simulation ($\sim$ 600 M$_\odot$) survives for a timescale of $\sim 10^5$ yr, leading to an injection rate of \HH\ from the CNM to the LNM of $\sim 10^{-9}$ M$_\odot$ yr$^{-1}$ per pc$^3$ of neutral diffuse ISM.

The natural follow up of this work is to extend the parametric study of the multiphase ISM and the methodology proposed for the comparison between simulations and observations to other chemical and kinematic tracers. Such investigations and their impact for the interpretation of future observations using the James Webb Space Telescope and the Square Kilometer Array are underway.

\begin{acknowledgements}

We are very grateful to the referee for their thorough reading and their clever comments which greatly improved the content of the paper. The research leading to these results has received fundings from the European Research Council, under the European Community's Seventh framework Programme, through the Advanced Grant MIST (FP7/2017-2022, No 742719). The grid of simulations used in this work has been run on the computing cluster Totoro of the ERC MIST. We would also like to acknowledge the support from the Programme National ``Physique et Chimie du Milieu Interstellaire'' (PCMI) of CNRS/INSU with INC/INP co-funded by CEA and CNES.

\end{acknowledgements}

\bibliographystyle{aa} 
\bibliography{mybib}

\appendix

\section{Observations of \CHp\ in the diffuse local ISM} \label{AppendObs}

The diffuse Galactic \CHp\ has been the target of many observational studies during the past 40 years. First limited to optical observations (e.g., \citealt{Federman1982,Sheffer2008}), these studies were recently extended to the infrared domain thanks to the {\it Herschel Space Telescope} (e.g., \citealt{Falgarone2010,Godard2012}) which allowed the observations of the submillimetre spectra of many hydrides previously unaccessible from the ground due to large atmospheric absorption lines \citep{Gerin2016}. In the optical, \CHp\ is observed in absorption against the continuum of bright nearby stars. The resulting lines of sight therefore sample nearby diffuse material over distances up to $\sim 2$ kpc (see Fig.~\ref{Fig-position-source}). In contrast, observations of \CHp\ in the submillimetre domain are performed in absorption against the infrared continuum of distant massive star forming regions located at small Galactic latitude. The resulting lines of sight therefore sample diffuse material across the entire Galactic disk over distances up to $\sim 12$ kpc. To avoid the complexity of modeling the Galactic structure and the necessary variations of physical conditions (e.g., the mean density of the diffuse gas, the density of OB stars, or the amount of mechanical energy injected at large scale) along extended lines of sight, we limit our present study to optical observations and refer to the sampled material as the Solar Neighborhood diffuse gas.

The observational sample studied in this paper is given in Table~\ref{Tab-Obs} which provides the Galactic coordinates of the sources, their distances, the total proton column density $N_{\rm H}$, and the column densities of \CHp. The column density of \CHp\ are taken from \citet{Federman1982, Lambert1986, Crane1995, Rachford2002, Gredel2002, Pan2004, Ritchey2006, Sheffer2008}. When available, all the other quantities listed in Table~\ref{Tab-Obs} are taken from the recent compilation of observation performed by \citet{Bellomi2020} (Table~A.1). If not available in this compilation, the quantities are derived as follows. The distances are computed from the parallaxes published in the {\it Gaia} EDR3 catalog \citep{Gaia-Collaboration2016,Gaia-Collaboration2021} or alternatively provided on the SIMBAD astronomical database \citep{Wenger2000}. The total proton column density $N_{\rm H}$ is derived from measurements of the reddening $E(B-V)$, as $N_{\rm H}=5.8 \times 10^{21} E(B-V)$ cm$^{-2}$ if such measurements are available. If not, $N_{\rm H}$ is derived from the Galactic dust extinction maps $A_V$, available on the NASA/IPAC infrared science archive \citep{Schlafly2011}, as $N_{\rm H}=1.87 \times 10^{21} A_V$ cm$^{-2}$. The relations between $N_{\rm H}$, $E(B-V)$, and $A_V$ are chosen assuming a standard Galactic extinction curve and the average interstellar ratio $R_V = A_V/E(B-V) = 3.1$ \citep{Fitzpatrick1986,Fitzpatrick1999}.

\onecolumn
\begingroup
\renewcommand{\arraystretch}{1.5} 
\begin{center}
\begin{longtable}{l @{\hspace{0.6cm}} 
                  r @{\hspace{0.6cm}} 
                  r @{\hspace{0.6cm}} 
                  r @{\hspace{0.6cm}} 
                  r @{\hspace{0.1cm}} 
                  l @{\hspace{0.0cm}} 
                  r @{\hspace{0.1cm}} 
                  l}
\caption{Observational dataset used in this work.}
\\
\hline
Source        &    longitude &   latitude   & distance & log$_{10}$$(N_{\rm H})$ & & log$_{10}$$(N(\CHp)$ & \\
              & ($^{\circ}$) & ($^{\circ}$) &    (kpc) &             (cm$^{-2}$) & &          (cm$^{-2}$) & \\
\hline
\endfirsthead
\caption{continued.} \\
\hline
Source        &    longitude &   latitude   & distance & log$_{10}$$(N_{\rm H})$ & & log$_{10}$$(N(\CHp)$ & \\
              & ($^{\circ}$) & ($^{\circ}$) &    (kpc) &             (cm$^{-2}$) & &          (cm$^{-2}$) & \\
\hline
\endhead
\hline
\endfoot
\hline
\hline
\multicolumn{8}{p{13cm}}{
{\bf Notes and References.} Observations of $N_{\rm H}$ are taken from the compilation of \citet{Bellomi2020} (Table~A.1) or computed from measurements of the reddening $E(B-V)$, as $N_{\rm H}=5.8 \times 10^{21} E(B-V)$ cm$^{-2}$, 
performed by 
$^{(1)} $\citet{Chaffee1982}, 
$^{(2)} $\citet{van-Dishoeck1989}, 
$^{(3)} $\citet{Federman1994},
$^{(4)} $\citet{Fruscione1994},
$^{(5)} $\citet{Diplas1994},
$^{(6)} $\citet{Crane1995},
$^{(7)} $\citet{Welsh1997},
$^{(8)} $\citet{Savage2001},
$^{(9)} $\citet{Cartledge2004},
$^{(10)}$\citet{Ritchey2006},
$^{(11)}$\citet{Bowen2008},
$^{(12)}$\citet{Sheffer2008}, and
$^{(13)}$\citet{Welty2010}. If no measurement of the reddening is found, $N_{\rm H}$ is derived from the Galactic dust extinction $A_V$, as $N_{\rm H}=1.87 \times 10^{21} A_V$ cm$^{-2}$, estimated by $^{(IR)}$\citet{Schlafly2011} and available on the NASA/IPAC infrared science archive.
Observations of $N(\CHp)$ are taken from 
$^{(a)} $\citet{Federman1982} (Table~1),
$^{(b)} $\citet{Lambert1986} (Table~1),
$^{(c)} $\citet{Crane1995} (Table~2),
$^{(d)} $\citet{Rachford2002} (Table~2),
$^{(e)} $\citet{Gredel2002} (Table~11)
$^{(f)} $\citet{Pan2004} (Table~6),
$^{(g)} $\citet{Ritchey2006} (Table~4), and
$^{(h)} $\citet{Sheffer2008} (Table~4).
}
\endlastfoot
  BD +48 3437 &  93.560 &  -2.060 &   2.639 &  21.45 &           & $ $ 13.18 & $^{(h)}$ \\
  BD +53 2820 & 101.240 &  -1.690 &   3.521 &  21.39 &           & $ $ 12.81 & $^{(h)}$ \\
 CPD -32 1734 & 248.160 &  -4.540 &   3.291 &  21.88 & $^{(IR)}$ & $ $ 13.78 & $^{(e)}$ \\
 CPD -33 1768 & 248.560 &  -4.100 &   2.002 &  21.93 & $^{(IR)}$ & $ $ 13.64 & $^{(e)}$ \\
 CPD -44 3129 & 264.690 &  -0.370 &   1.993 &  22.17 & $^{(IR)}$ & $ $ 13.80 & $^{(e)}$ \\
 CPD -45 3218 & 266.180 &  -0.850 &   1.910 &  22.03 & $^{(IR)}$ & $ $ 13.45 & $^{(e)}$ \\
 CPD -46 3272 & 267.350 &  -1.030 &   2.201 &  21.76 & $^{(IR)}$ & $ $ 13.51 & $^{(e)}$ \\
 CPD -59 2603 & 287.590 &  -0.690 &   4.098 &  21.50 &           & $ $ 13.20 & $^{(h)}$ \\
 CPD -59 4551 & 303.220 &  +2.540 &   2.119 &  21.86 & $^{(IR)}$ & $ $ 13.43 & $^{(e)}$ \\
 CPD -69 1743 & 303.710 &  -7.350 &   3.817 &  21.18 &           & $ $ 13.18 & $^{(h)}$ \\
    HD 002905 & 120.840 &  +0.140 &   0.521 &  21.29 &           & $ $ 12.56 & $^{(a)}$ \\
    HD 012323 & 132.910 &  -5.870 &   2.809 &  21.29 &           & $ $ 12.90 & $^{(h)}$ \\
    HD 013268 & 133.960 &  -4.990 &   1.692 &  21.45 &           & $ $ 13.18 & $^{(h)}$ \\
    HD 013745 & 134.580 &  -4.960 &   2.268 &  21.44 &           & $ $ 13.52 & $^{(h)}$ \\
    HD 014434 & 135.080 &  -3.820 &   2.558 &  21.53 &           & $ $ 13.38 & $^{(h)}$ \\
    HD 015137 & 137.460 &  -7.580 &   3.704 &  21.23 &           & $ $ 13.15 & $^{(h)}$ \\
    HD 021278 & 147.520 &  -6.190 &   0.178 &  21.29 &           & $ $ 12.56 & $^{(a)}$ \\
    HD 022951 & 158.920 & -16.703 &   0.370 &  21.36 & $^{(IR)}$ & $ $ 12.84 & $^{(c)}$ \\
    HD 023180 & 160.360 & -17.740 &   0.245 &  21.21 &           & $ $ 12.84 & $^{(h)}$ \\
    HD 023288 & 166.040 & -23.730 &   0.135 &  20.76 & $^{(10)}$ & $ $ 13.34 & $^{(g)}$ \\
    HD 023302 & 166.180 & -23.850 &   0.120 &  20.46 & $^{(10)}$ & $ $ 12.34 & $^{(g)}$ \\
    HD 023324 & 165.710 & -23.260 &   0.138 &  20.46 & $^{(10)}$ & $ $ 12.56 & $^{(g)}$ \\
    HD 023338 & 165.980 & -23.530 &   0.105 &  20.37 & $^{(10)}$ & $ $ 12.52 & $^{(g)}$ \\
    HD 023408 & 166.170 & -23.510 &   0.130 &  20.61 &           & $ $ 13.50 & $^{(g)}$ \\
    HD 023410 & 167.070 & -24.420 &   0.135 &  20.67 & $^{(10)}$ & $ $ 12.43 & $^{(g)}$ \\
    HD 023432 & 166.050 & -23.360 &   0.137 &  20.61 & $^{(10)}$ & $ $ 13.37 & $^{(g)}$ \\
    HD 023441 & 166.090 & -23.360 &   0.135 &  20.54 & $^{(10)}$ & $ $ 13.27 & $^{(g)}$ \\
    HD 023478 & 160.760 & -17.420 &   0.288 &  21.21 &           & $ $ 12.32 & $^{(h)}$ \\
    HD 023480 & 166.570 & -23.750 &   0.106 &  20.76 &           & $ $ 13.29 & $^{(g)}$ \\
    HD 023512 & 166.850 & -23.950 &   0.136 &  21.31 & $^{(10)}$ & $ $ 12.79 & $^{(g)}$ \\
    HD 023568 & 166.270 & -23.220 &   0.138 &  20.61 & $^{(10)}$ & $ $ 13.41 & $^{(g)}$ \\
    HD 023629 & 166.640 & -23.470 &   0.139 &  21.07 & $^{(IR)}$ & $ $ 12.40 & $^{(g)}$ \\
    HD 023630 & 166.670 & -23.460 &   0.125 &  20.28 &           & $ $ 12.38 & $^{(g)}$ \\
    HD 023753 & 167.330 & -23.830 &   0.130 &  20.37 & $^{(10)}$ & $ $ 12.62 & $^{(g)}$ \\
    HD 023850 & 167.010 & -23.230 &   0.123 &  20.37 & $^{(10)}$ & $ $ 12.69 & $^{(g)}$ \\
    HD 023862 & 166.960 & -23.170 &   0.138 &  20.24 & $^{(10)}$ & $ $ 12.69 & $^{(g)}$ \\
    HD 023873 & 166.810 & -22.960 &   0.139 &  20.06 & $^{(10)}$ & $ $ 12.64 & $^{(g)}$ \\
    HD 023923 & 167.370 & -23.400 &   0.134 &  20.54 & $^{(10)}$ & $ $ 12.45 & $^{(g)}$ \\
    HD 024076 & 167.390 & -23.040 &   0.147 &  20.24 & $^{(10)}$ & $ $ 12.67 & $^{(g)}$ \\
    HD 024190 & 160.390 & -15.180 &   0.413 &  21.30 &           & $ $ 13.18 & $^{(h)}$ \\
    HD 024398 & 162.290 & -16.690 &   0.294 &  21.20 &           & $ $ 12.45 & $^{(h)}$ \\
    HD 024534 & 163.080 & -17.140 &   0.810 &  21.34 &           & $ $ 13.22 & $^{(d)}$ \\
    HD 024912 & 160.370 & -13.110 &   0.725 &  21.29 &           & $ $ 13.47 & $^{(c)}$ \\
    HD 027778 & 172.760 & -17.390 &   0.224 &  21.40 &           & $ $ 12.85 & $^{(d)}$ \\
    HD 030122 & 176.620 & -14.030 &   0.257 &  21.54 &           & $ $ 12.48 & $^{(h)}$ \\
    HD 030614 & 144.070 &  14.040 &   0.730 &  21.09 &           & $ $ 13.29 & $^{(c)}$ \\
    HD 034078 & 172.080 &  -2.260 &   0.406 &  21.55 &           & $ $ 13.84 & $^{(h)}$ \\
    HD 035149 & 199.160 & -17.860 &   0.368 &  20.74 &           & $ $ 13.01 & $^{(c)}$ \\
    HD 035411 & 204.866 & -20.392 &   0.634 &  21.07 & $^{(IR)}$ & $ $ 12.14 & $^{(b)}$ \\
    HD 036841 & 204.260 & -17.220 &   0.418 &  21.29 & $^{(4)}$  & $ $ 12.76 & $^{(h)}$ \\
    HD 037367 & 179.040 &  -1.030 &   0.989 &  21.43 &           & $ $ 13.51 & $^{(h)}$ \\
    HD 037903 & 206.850 & -16.540 &   0.399 &  21.46 &           & $ $ 13.11 & $^{(h)}$ \\
    HD 041117 & 189.650 &  -0.860 &   1.309 &  21.54 &           & $ $ 13.38 & $^{(c)}$ \\
    HD 043818 & 188.490 &  +3.870 &   2.147 &  21.53 & $^{(5)}$  & $ $ 13.18 & $^{(h)}$ \\
    HD 052382 & 222.170 &  -2.150 &   2.476 &  21.98 & $^{(IR)}$ & $ $ 13.38 & $^{(e)}$ \\
    HD 053755 & 224.050 &  -1.690 &   1.100 &  21.13 & $^{(4)}$  & $ $ 12.95 & $^{(e)}$ \\
    HD 053975 & 225.680 &  -2.320 &   1.247 &  21.16 &           & $ $ 12.48 & $^{(e)}$ \\
    HD 054662 & 224.170 &  -0.780 &   1.170 &  21.41 &           & $ $ 13.04 & $^{(e)}$ \\
    HD 055879 & 224.730 &  +0.350 &   1.011 &  20.84 &           & $ $ 12.54 & $^{(e)}$ \\
    HD 058510 & 235.520 &  -2.470 &   3.333 &  21.31 &           & $ $ 13.08 & $^{(h)}$ \\
    HD 061827 & 247.120 &  -5.070 &   4.560 &  21.55 & $^{(IR)}$ & $ $ 13.11 & $^{(e)}$ \\
    HD 062150 & 247.320 &  -4.840 &   4.160 &  21.59 & $^{(IR)}$ & $ $ 13.46 & $^{(e)}$ \\
    HD 062844 & 247.320 &  -4.030 &   3.520 &  21.80 & $^{(IR)}$ & $ $ 13.51 & $^{(e)}$ \\
    HD 063005 & 242.470 &  -0.930 &  13.699 &  21.32 &           & $ $ 13.15 & $^{(h)}$ \\
    HD 063423 & 246.170 &  -2.630 &   1.520 &  21.61 & $^{(IR)}$ & $ $ 13.04 & $^{(e)}$ \\
    HD 063804 & 248.770 &  -3.710 &   3.437 &  21.84 & $^{(2)}$  & $ $ 13.68 & $^{(e)}$ \\
    HD 073882 & 260.180 &  +0.640 &   0.347 &  21.59 &           & $ $ 13.38 & $^{(e)}$ \\
    HD 074194 & 264.040 &  -1.950 &   2.361 &  21.45 & $^{(8)}$  & $ $ 13.23 & $^{(e)}$ \\
    HD 074371 & 264.440 &  -2.010 &   1.669 &  21.89 & $^{(IR)}$ & $ $ 13.15 & $^{(e)}$ \\
    HD 075149 & 265.330 &  -1.690 &   1.455 &  21.38 & $^{(2)}$  & $ $ 13.08 & $^{(e)}$ \\
    HD 075211 & 263.960 &  -0.470 &   1.580 &  22.06 & $^{(IR)}$ & $ $ 13.57 & $^{(e)}$ \\
    HD 075860 & 264.140 &  +0.270 &   2.198 &  22.22 & $^{(IR)}$ & $ $ 13.88 & $^{(e)}$ \\
    HD 076556 & 267.580 &  -1.630 &   1.870 &  21.42 & $^{(IR)}$ & $ $ 13.18 & $^{(e)}$ \\
    HD 078344 & 268.890 &  -0.380 &   2.213 &  21.91 & $^{(2)}$  & $ $ 13.54 & $^{(e)}$ \\
    HD 091983 & 285.880 &  +0.050 &   4.255 &  21.24 &           & $ $ 12.52 & $^{(h)}$ \\
    HD 093840 & 282.140 &  11.100 &   3.521 &  21.05 &           & $ $ 12.52 & $^{(h)}$ \\
    HD 096675 & 296.620 & -14.570 &   0.163 &  21.25 &           & $ $ 13.45 & $^{(d)}$ \\
    HD 099872 & 296.690 & -10.620 &   0.230 &  21.32 &           & $ $ 13.36 & $^{(h)}$ \\
    HD 100262 & 292.980 &  +1.810 &   2.143 &  21.66 & $^{(IR)}$ & $ $ 12.89 & $^{(b)}$ \\
    HD 102065 & 300.030 & -18.000 &   0.194 &  20.99 &           & $ $ 13.04 & $^{(h)}$ \\
    HD 110432 & 301.960 &  -0.200 &   0.420 &  21.20 &           & $ $ 13.25 & $^{(d)}$ \\
    HD 110639 & 302.080 &  +1.470 &   2.791 &  22.05 & $^{(IR)}$ & $ $ 13.56 & $^{(e)}$ \\
    HD 111904 & 303.170 &  +2.540 &   2.254 &  21.87 & $^{(IR)}$ & $ $ 13.08 & $^{(e)}$ \\
    HD 111934 & 303.200 &  +2.510 &   2.058 &  21.27 & $^{(9)}$  & $ $ 13.08 & $^{(e)}$ \\
    HD 111973 & 303.000 &  +2.500 &   1.976 &  21.19 & $^{(7)}$  & $ $ 13.11 & $^{(e)}$ \\
    HD 111990 & 303.250 &  +2.530 &   2.482 &  21.86 & $^{(IR)}$ & $ $ 13.18 & $^{(e)}$ \\
    HD 112244 & 303.550 &  +6.030 &   1.167 &  21.19 &           & $ $ 12.88 & $^{(b)}$ \\
    HD 112272 & 303.490 &  -1.490 &   1.600 &  21.78 & $^{(2)}$  & $ $ 13.58 & $^{(e)}$ \\
    HD 112366 & 303.570 &  -0.600 &   2.061 &  22.82 & $^{(IR)}$ & $ $ 13.32 & $^{(e)}$ \\
    HD 113422 & 304.490 &  +1.120 &   1.175 &  21.77 & $^{(2)}$  & $ $ 13.75 & $^{(e)}$ \\
    HD 113432 & 304.420 &  -0.660 &   2.037 &  22.53 & $^{(IR)}$ & $ $ 13.57 & $^{(e)}$ \\
    HD 113904 & 304.670 &  -2.490 &   2.786 &  21.13 &           & $ $ 12.35 & $^{(b)}$ \\
    HD 114011 & 305.010 &  +1.620 &   3.088 &  22.09 & $^{(IR)}$ & $ $ 13.90 & $^{(e)}$ \\
    HD 114213 & 305.190 &  +1.320 &   1.760 &  21.81 & $^{(13)}$ & $ $ 13.11 & $^{(e)}$ \\
    HD 114886 & 305.520 &  -0.830 &   1.045 &  21.42 &           & $ $ 13.28 & $^{(h)}$ \\
    HD 115363 & 305.880 &  -0.970 &   2.881 &  22.12 & $^{(IR)}$ & $ $ 13.26 & $^{(e)}$ \\
    HD 115455 & 306.060 &  +0.220 &   2.268 &  21.52 &           & $ $ 13.23 & $^{(h)}$ \\
    HD 115704 & 306.300 &  +0.680 &   3.420 &  21.81 &           & $ $ 13.45 & $^{(e)}$ \\
    HD 116852 & 304.880 & -16.130 &  22.727 &  21.02 &           & $ $ 12.54 & $^{(h)}$ \\
    HD 122879 & 312.260 &  +1.790 &   2.387 &  21.35 &           & $ $ 13.08 & $^{(h)}$ \\
    HD 124314 & 312.670 &  -0.420 &   1.808 &  21.49 &           & $ $ 13.18 & $^{(h)}$ \\
    HD 137595 & 336.720 & +18.860 &   0.822 &  21.24 &           & $ $ 13.26 & $^{(h)}$ \\
    HD 143018 & 347.210 & +20.230 &   0.580 &  20.77 &           & $ $ 11.76 & $^{(b)}$ \\
    HD 143275 & 350.100 & +22.490 &   0.155 &  21.17 &           & $ $ 12.26 & $^{(c)}$ \\
    HD 144217 & 353.190 & +23.600 &   0.161 &  21.13 &           & $ $ 12.71 & $^{(c)}$ \\
    HD 144218 & 353.200 & +23.600 &   0.121 &  21.12 & $^{(IR)}$ & $ $ 13.01 & $^{(b)}$ \\
    HD 144470 & 352.750 & +22.770 &   0.142 &  21.24 &           & $ $ 12.81 & $^{(b)}$ \\
    HD 144965 & 339.040 &  +8.420 &   0.266 &  21.37 &           & $ $ 12.88 & $^{(h)}$ \\
    HD 145502 & 354.610 & +22.700 &   0.135 &  21.20 &           & $ $ 12.82 & $^{(b)}$ \\
    HD 147165 & 351.310 & +17.000 &   0.100 &  21.40 &           & $ $ 12.77 & $^{(b)}$ \\
    HD 147683 & 344.860 & +10.090 &   0.295 &  21.55 &           & $ $ 13.28 & $^{(h)}$ \\
    HD 147888 & 353.650 & +17.710 &   0.092 &  21.77 &           & $ $ 12.88 & $^{(h)}$ \\
    HD 147932 & 353.720 & +17.710 &   0.134 &  21.44 & $^{(3)}$  & $ $ 12.85 & $^{(f)}$ \\
    HD 147933 & 353.690 & +17.690 &   0.174 &  21.86 &           & $ $ 13.20 & $^{(f)}$ \\
    HD 147934 & 353.690 & +17.690 &   0.182 &  21.78 & $^{(IR)}$ & $ $ 13.16 & $^{(f)}$ \\
    HD 148184 & 357.930 & +20.680 &   0.122 &  21.36 &           & $ $ 13.14 & $^{(c)}$ \\
    HD 148605 & 353.100 & +15.800 &   0.117 &  20.96 &           & $ $ 12.10 & $^{(b)}$ \\
    HD 149038 & 339.380 &  +2.510 &   0.842 &  21.27 &           & $ $ 13.60 & $^{(b)}$ \\
    HD 149757 &   6.280 & +23.590 &   0.172 &  21.17 &           & $ $ 13.45 & $^{(c)}$ \\
    HD 151804 & 343.620 &  +1.940 &   1.629 &  21.19 &           & $ $ 12.92 & $^{(b)}$ \\
    HD 152236 & 343.030 &  +0.870 &   1.403 &  21.84 &           & $ $ 13.24 & $^{(b)}$ \\
    HD 152590 & 344.840 &  +1.830 &   1.637 &  21.47 &           & $ $ 13.28 & $^{(h)}$ \\
    HD 152723 & 344.810 &  +1.610 &  16.667 &  21.49 &           & $ $ 13.04 & $^{(h)}$ \\
    HD 154090 & 350.830 &  +4.290 &   1.083 &  21.42 & $^{(5)}$  & $ $ 13.33 & $^{(b)}$ \\
    HD 154368 & 349.970 &  +3.220 &   1.217 &  21.59 &           & $ $ 13.67 & $^{(d)}$ \\
    HD 155806 & 352.590 &  +2.870 &   0.994 &  21.14 &           & $ $ 12.83 & $^{(b)}$ \\
    HD 157246 & 334.640 & -11.480 &   0.267 &  20.77 &           & $ $ 11.91 & $^{(b)}$ \\
    HD 157857 &  12.970 & +13.310 &   3.968 &  21.47 &           & $ $ 13.30 & $^{(h)}$ \\
    HD 159975 &  17.000 & +12.340 &   0.200 &  21.62 & $^{(IR)}$ & $ $ 13.07 & $^{(b)}$ \\
    HD 163758 & 355.360 &  -6.100 &   3.876 &  21.26 &           & $ $ 12.15 & $^{(h)}$ \\
    HD 164353 &  29.730 & +12.630 &   0.566 &  21.13 &           & $ $ 12.71 & $^{(c)}$ \\
    HD 166937 &  10.000 &  -1.600 &   1.451 &  21.14 & $^{(6)}$  & $ $ 13.05 & $^{(c)}$ \\
    HD 167263 &  10.760 &  -1.580 &   2.079 &  21.18 &           & $ $ 12.78 & $^{(b)}$ \\
    HD 167264 &  10.460 &  -1.740 &   1.140 &  21.25 &           & $ $ 12.91 & $^{(b)}$ \\
    HD 167971 &  18.250 &  +1.680 &   2.033 &  21.73 &           & $ $ 13.73 & $^{(d)}$ \\
    HD 169454 &  17.540 &  -0.670 &   2.128 &  21.81 &           & $ $ 12.14 & $^{(b)}$ \\
    HD 170740 &  21.060 &  -0.530 &   0.231 &  21.46 &           & $ $ 13.26 & $^{(d)}$ \\
    HD 184915 &  31.770 & -13.290 &   0.466 &  21.05 &           & $ $ 12.78 & $^{(c)}$ \\
    HD 185418 &  53.600 &  -2.170 &   0.755 &  21.39 &           & $ $ 13.10 & $^{(d)}$ \\
    HD 190918 &  72.650 &  +2.070 &   1.953 &  21.43 &           & $ $ 13.15 & $^{(h)}$ \\
    HD 192035 &  83.330 &  +7.760 &   2.252 &  21.41 &           & $ $ 12.89 & $^{(h)}$ \\
    HD 192639 &  74.900 &  +1.480 &   2.597 &  21.49 &           & $ $ 13.61 & $^{(h)}$ \\
    HD 198781 &  99.940 & +12.610 &   0.935 &  21.15 &           & $ $ 12.52 & $^{(h)}$ \\
    HD 199579 &  85.700 &  -0.300 &   0.941 &  21.25 &           & $ $ 13.01 & $^{(d)}$ \\
    HD 200775 & 104.060 &  14.190 &   0.361 &  21.52 &           & $ $ 12.97 & $^{(h)}$ \\
    HD 203064 &  87.610 &  -3.840 &   0.587 &  21.14 &           & $ $ 12.89 & $^{(c)}$ \\
    HD 203374 & 100.510 &  +8.620 &   2.611 &  21.38 &           & $ $ 12.85 & $^{(f)}$ \\
    HD 203532 & 309.460 & -31.740 &   0.292 &  21.44 &           & $ $ 12.48 & $^{(h)}$ \\
    HD 203938 &  90.560 &  -2.330 &   0.223 &  21.70 &           & $ $ 13.68 & $^{(d)}$ \\
    HD 204827 &  99.170 &  +5.550 &   0.929 &  21.72 & $^{(IR)}$ & $ $ 13.57 & $^{(f)}$ \\
    HD 206165 & 102.270 &  +7.250 &   0.746 &  21.44 &           & $ $ 13.20 & $^{(f)}$ \\
    HD 206183 &  98.890 &  +3.400 &   0.921 &  21.86 & $^{(IR)}$ & $ $ 13.18 & $^{(f)}$ \\
    HD 206267 &  99.290 &  +3.740 &   1.117 &  21.54 &           & $ $ 13.02 & $^{(d)}$ \\
    HD 206773 &  99.800 &  +3.620 &   0.958 &  21.25 &           & $ $ 13.17 & $^{(f)}$ \\
    HD 207198 & 103.140 &  +6.990 &   1.025 &  21.55 &           & $ $ 13.34 & $^{(f)}$ \\
    HD 207260 & 102.310 &  +5.930 &   1.073 &  21.42 & $^{(1)}$  & $ $ 13.28 & $^{(f)}$ \\
    HD 207308 & 103.110 &  +6.820 &   1.026 &  21.44 &           & $ $ 13.22 & $^{(f)}$ \\
    HD 207538 & 101.600 &  +4.670 &   0.838 &  21.58 &           & $ $ 12.94 & $^{(f)}$ \\
    HD 208266 & 102.710 &  +4.980 &   0.911 &  21.48 &           & $ $ 12.71 & $^{(f)}$ \\
    HD 208440 & 104.030 &  +6.440 &   0.829 &  21.33 &           & $ $ 12.94 & $^{(f)}$ \\
    HD 208501 & 100.390 &  +1.680 &   1.096 &  21.65 & $^{(1)}$  & $ $ 12.56 & $^{(f)}$ \\
    HD 208905 & 103.530 &  +5.170 &   1.031 &  21.33 & $^{(12)}$ & $ $ 12.78 & $^{(h)}$ \\
    HD 209339 & 104.580 &  +5.870 &   0.845 &  21.25 &           & $ $ 12.82 & $^{(f)}$ \\
    HD 209481 & 101.010 &  +2.180 &   1.101 &  21.30 &           & $ $ 12.72 & $^{(h)}$ \\
    HD 209975 & 104.870 &  +5.390 &   0.858 &  21.20 &           & $ $ 13.38 & $^{(h)}$ \\
    HD 210121 &  56.880 & -44.460 &   0.342 &  21.19 &           & $ $ 12.78 & $^{(d)}$ \\
    HD 210809 &  99.850 &  -3.130 &   4.329 &  21.30 &           & $ $ 12.88 & $^{(h)}$ \\
    HD 210839 & 103.830 &  +2.610 &   0.617 &  21.43 &           & $ $ 13.17 & $^{(c)}$ \\
    HD 216532 & 109.650 &  +2.680 &   0.751 &  21.70 &           & $ $ 13.65 & $^{(f)}$ \\
    HD 216898 & 109.930 &  +2.390 &   0.840 &  21.69 &           & $ $ 13.42 & $^{(f)}$ \\
    HD 217035 & 110.250 &  +2.860 &   0.829 &  21.67 &           & $ $ 13.32 & $^{(f)}$ \\
    HD 217312 & 110.560 &  +2.950 &   1.631 &  21.63 &           & $ $ 13.67 & $^{(f)}$ \\
    HD 218376 & 109.950 &  -0.780 &   0.374 &  21.07 &           & $ $ 12.91 & $^{(a)}$ \\
    HD 220057 & 112.130 &  +0.210 &   0.392 &  21.27 &           & $ $ 12.87 & $^{(h)}$ \\
    HD 224572 & 115.550 &  -6.360 &   0.292 &  21.04 &           & $ $ 12.75 & $^{(a)}$ \\
    HD 000886 & 109.434 & -46.684 &   0.144 &  20.46 & $^{(IR)}$ & $<$ 12.08 & $^{(a)}$ \\
    HD 010144 & 290.840 & -58.790 &   0.043 &  20.88 & $^{(11)}$ & $<$ 11.10 & $^{(b)}$ \\
    HD 014228 & 275.350 & -60.820 &   0.046 &  20.17 & $^{(IR)}$ & $<$ 11.70 & $^{(b)}$ \\
    HD 022828 & 131.647 & +18.367 &   0.266 &  20.92 & $^{(IR)}$ & $<$ 12.11 & $^{(a)}$ \\
    HD 023964 & 167.310 & -23.260 &   0.136 &  20.80 & $^{(10)}$ & $<$ 12.87 & $^{(g)}$ \\
    HD 024760 & 157.354 & -10.088 &   0.186 &  20.54 &           & $<$ 11.90 & $^{(a)}$ \\
    HD 035039 & 202.630 & -20.030 &   0.350 &  20.92 & $^{(IR)}$ & $<$ 11.80 & $^{(b)}$ \\
    HD 036486 & 203.856 & -17.740 &   0.212 &  20.18 &           & $<$ 12.15 & $^{(a)}$ \\
    HD 036822 & 195.400 & -12.290 &   0.348 &  20.84 &           & $<$ 12.00 & $^{(b)}$ \\
    HD 036861 & 195.052 & -11.995 &   0.405 &  20.89 &           & $<$ 12.11 & $^{(a)}$ \\
    HD 037043 & 209.522 & -19.583 &   0.501 &  20.30 &           & $<$ 12.04 & $^{(a)}$ \\
    HD 037128 & 205.212 & -17.242 &   0.606 &  20.45 &           & $<$ 11.94 & $^{(a)}$ \\
    HD 037202 & 185.686 & -05.636 &   0.136 &  21.39 & $^{(IR)}$ & $<$ 12.28 & $^{(a)}$ \\
    HD 038771 & 214.514 & -18.496 &   0.198 &  20.52 &           & $<$ 12.15 & $^{(a)}$ \\
    HD 044743 & 226.060 & -14.268 &   0.151 &  20.97 & $^{(IR)}$ & $<$ 12.04 & $^{(a)}$ \\
    HD 047839 & 202.936 & +02.198 &   0.282 &  20.31 &           & $<$ 12.48 & $^{(a)}$ \\
    HD 052089 & 239.830 & -11.330 &   0.132 &  19.76 &           & $<$ 11.90 & $^{(b)}$ \\
    HD 057061 & 238.180 &  -5.540 &   1.735 &  20.74 &           & $<$ 11.70 & $^{(b)}$ \\
    HD 062542 & 255.920 &  -9.240 &   0.390 &  21.32 &           & $<$ 11.83 & $^{(d)}$ \\
    HD 074575 & 254.990 &  +5.770 &   0.235 &  20.61 &           & $<$ 11.40 & $^{(b)}$ \\
    HD 087901 & 226.427 & +48.934 &   0.024 &  20.26 & $^{(IR)}$ & $<$ 11.90 & $^{(a)}$ \\
    HD 091316 & 234.890 & +52.770 &   0.505 &  20.26 &           & $<$ 11.50 & $^{(b)}$ \\
    HD 093030 & 289.600 &  -4.900 &   0.131 &  20.54 &           & $<$ 11.10 & $^{(b)}$ \\
    HD 106490 & 298.230 &  +3.790 &   0.086 &  20.06 &           & $<$ 11.80 & $^{(b)}$ \\
    HD 108642 & 221.680 & +84.830 &   0.086 &  20.04 & $^{(IR)}$ & $<$ 11.40 & $^{(b)}$ \\
    HD 116658 & 316.112 & +50.845 &   0.077 &  20.48 & $^{(IR)}$ & $<$ 11.85 & $^{(a)}$ \\
    HD 118716 & 310.190 &  +8.720 &   0.168 &  20.37 &           & $<$ 11.40 & $^{(b)}$ \\
    HD 120307 & 314.410 & +19.890 &   0.124 &  19.76 & $^{(11)}$ & $<$ 11.40 & $^{(b)}$ \\
    HD 120315 & 100.696 & +65.322 &   0.032 &  19.76 & $^{(IR)}$ & $<$ 11.79 & $^{(a)}$ \\
    HD 121263 & 314.070 & +14.190 &   0.120 &  19.28 &           & $<$ 11.80 & $^{(b)}$ \\
    HD 121743 & 315.980 & +19.070 &   0.141 &  20.64 & $^{(IR)}$ & $<$ 11.40 & $^{(b)}$ \\
    HD 132058 & 326.250 & +13.910 &   0.117 &  20.24 & $^{(11)}$ & $<$ 11.50 & $^{(b)}$ \\
    HD 135742 & 352.020 & +39.230 &   0.093 &  19.38 &           & $<$ 11.20 & $^{(b)}$ \\
    HD 136298 & 331.320 & +13.820 &   0.115 &  20.06 &           & $<$ 11.40 & $^{(b)}$ \\
    HD 138690 & 333.190 & +11.890 &   0.129 &  20.24 &           & $<$ 11.20 & $^{(b)}$ \\
    HD 141637 & 346.100 & +21.710 &   0.145 &  21.19 &           & $<$ 12.10 & $^{(b)}$ \\
    HD 143118 & 338.774 & +11.009 &   0.132 &  21.29 & $^{(IR)}$ & $<$ 11.50 & $^{(b)}$ \\
    HD 149438 & 351.530 & +12.810 &   0.195 &  20.43 &           & $<$ 11.10 & $^{(b)}$ \\
    HD 151890 & 346.120 &  +3.910 &   0.268 &  20.46 &           & $<$ 11.70 & $^{(b)}$ \\
    HD 151995 &  37.859 & +34.959 &   0.027 &  20.59 & $^{(IR)}$ & $<$ 11.40 & $^{(b)}$ \\
    HD 157056 &   0.460 &  +6.550 &   0.102 &  20.24 & $^{(4)}$  & $<$ 12.00 & $^{(b)}$ \\
    HD 158408 & 351.270 &  -1.840 &   0.134 &  20.06 &           & $<$ 11.50 & $^{(b)}$ \\
    HD 158926 & 351.740 &  -2.210 &   0.220 &  19.23 &           & $<$ 11.40 & $^{(b)}$ \\
    HD 160578 & 301.040 &  -4.720 &   0.202 &  20.68 &           & $<$ 11.40 & $^{(b)}$ \\
    HD 165024 & 343.330 & -13.820 &   0.279 &  20.86 &           & $<$ 11.80 & $^{(b)}$ \\
    HD 175191 &   9.560 & -12.440 &   0.070 &  20.46 &           & $<$ 11.20 & $^{(b)}$ \\
    HD 187642 &  47.740 &  -8.910 &   0.005 &  20.24 & $^{(4)}$  & $<$ 11.40 & $^{(b)}$ \\
    HD 193924 & 340.900 & -35.190 &   0.056 &  20.06 &           & $<$ 11.20 & $^{(b)}$ \\
    HD 200120 & 088.030 & +00.971 &   0.681 &  20.34 &           & $<$ 12.48 & $^{(a)}$ \\
    HD 207971 &   6.110 & -51.470 &   0.056 &  20.06 & $^{(IR)}$ & $<$ 11.40 & $^{(b)}$ \\
    HD 209952 & 350.000 & -52.470 &   0.030 &  20.54 &           & $<$ 11.10 & $^{(b)}$ \\
    HD 210191 &  37.151 & -51.762 &   0.749 &  20.09 & $^{(IR)}$ & $<$ 11.50 & $^{(b)}$ \\
    HD 214680 & 096.651 & -16.983 &   0.456 &  20.73 &           & $<$ 12.36 & $^{(a)}$ \\
    HD 217675 & 102.208 & -16.096 &   0.107 &  21.09 & $^{(IR)}$ & $<$ 11.85 & $^{(a)}$ \\
\label{Tab-Obs}
\end{longtable}
\end{center}
\endgroup

\end{document}